\documentclass[12pt]{article}

\usepackage{a4,latexsym,pslatex,graphicx,times,color,amsmath}
\usepackage{subcaption}

\usepackage[numbers]{natbib}

\newcommand{\dd}[1]{{\rm{d}}{#1}}

\title{Slip flows of a Bingham fluid in curved channels}
\author{S.J. Cox$\;^1$ and S.M. Taghavi$\;^2$\\ 
$\,^1$Department of Mathematics, Aberystwyth University, SY23 3BZ, UK.\\
$\,^2$ Department of Chemical Engineering, Universit{\'e} Laval, QC G1V 0A6, Canada
}
\date{}

\begin{document}

\maketitle

\begin{abstract}
{Motivated by
the treatment of varicose veins with aqueous foams, 
we determine the velocity profiles and yield surface positions for pressure-driven flows of a Bingham fluid in curved channels, incorporating Navier wall slip to capture the slip behavior of aqueous foams}. Assuming a constant pressure gradient -- consistent with syringe-driven injection in sclerotherapy -- we reproduce a closed-form analytic solution for the flow in a straight channel, and extend this to a uniformly curved channel consisting of a section of an annulus. We find that increasing the slip length in the curved channel moves the unyielded fluid, or plug, towards the inside of the channel and slightly decreases the plug width. Channels with a change in curvature are tackled numerically, with a procedure that is validated against the analytic results. We solve for the flow of a Bingham fluid in the transition region between a straight and a uniformly curved channel and in a non-uniform sinusoidal channel  where the wall curvature changes continuously. We find that there is greater yielding of the fluid in curved channels, caused by an increase in the shear stresses. As the slip length increases, there is a  reduction in the area of unyielded fluid in regions of high wall curvature. In the context of varicose vein sclerotherapy, our results suggest that slip has both positive and negative consequences: it reduces the possibility of static regions of foam developing in areas of high curvature in the vein, but also reduces the plug width in general, reducing the efficacy with which the foam displaces blood from the vein.
\end{abstract}

\section{Introduction}

{Aqueous foams---comprising gas bubbles dispersed in a liquid phase---are a type of of yield stress fluids that are characterized by their ability to resist deformation under small applied stresses due to a jammed, disordered microstructure of closely packed bubbles~\cite{hohlerca05,mousse13}. In such materials, flow occurs only when the applied shear stress exceeds a critical value, called the yield stress $\tau_y$, which depends on the liquid fraction $\phi_l$ (or equivalently the gas fraction $\phi = 1 - \phi_l$); as $\phi_l$ increases, bubble rearrangement becomes easier, reducing $\tau_y$. This interesting rheological behavior is also common to other complex fluids such as emulsions (e.g. paint, mayonnaise) and pastes (e.g. toothpaste), where the microstructure also governs macroscopic flow properties. In foams, however, the microstructure is dominated by bubble interface interactions, and the rheology is tied to bubble deformation. In this context, our specific motivation for this work comes from the use of aqueous foams in the treatment of varicose veins, where a foam is injected to displace stagnant blood and deliver a surfactant (sclerosant) to the vein walls, triggering their collapse~\cite{eckmann9,nastasa15,robertsclj21}. The yield stress plays a central role in this procedure by minimizing mixing between blood and foam, thereby maintaining the efficacy of the displacement. Since varicose veins can be up to 5 mm in diameter~\cite{meghdadijplmc21} and often exhibit bulges and curvature due to blood pooling, understanding foam flow in curved geometries is essential to optimizing treatment outcomes.}

In the process of foam sclerotherapy, it is important to optimize the foam parameters, such as bubble size and liquid fraction, and its injection rate. The yield stress not only depends on the liquid fraction, but also the bubble size -- smaller bubbles lead to a greater yield stress~\cite{princen83}. A large yield stress is good for the displacement flow, but it increases the applied pressure required to inject the foam into the vein, possibly making delivery of the foam into the vein too difficult. 

Yield stress fluids are often represented using the Hershel-Bulkley constitutive model, in which the fluid behaves as a power-law fluid above the yield stress and is jammed (static) below it. Elaborate extensions to this model can include elasticity~\cite{saramito07,saramito09} and thixotropy~\cite{varchanismmdt19}, but it remains challenging to relate macroscopic parameters, such as the effective viscosity and yield stress, to local quantities defined on the bubble-scale, such as the bubble size distribution, the degree of bubble deformation, and viscous dissipation in the liquid~\cite{kraynik88,denkovsgl05}. This remains true even in its most simple incarnation, the Bingham model~\cite{bingham}, in which the power-law exponent is one and the fluid responds linearly to the strain-rate above the yield stress, as for a Newtonian fluid.

There are a few channel geometries for which closed-form solutions to the Stokes equations for slow fluid motion can be solved with the Bingham constitutive model~\cite{birddy83}, in both 2D and 3D, but more generally numerical solutions are required~\cite{abdali_mitsoulis, blackery_mitsoulis,mitsoulis, muravleva_mitsoulis}. Roberts and Cox~\cite{robertsc20} derived a closed form expression for the velocity field and plug width in the pressure-driven flow of a Bingham fluid in a curved 2D channel formed from a section of an annulus with constant width and curvature. They assumed that the walls of the channel had a no-slip condition; we extend that work here to include a more general boundary condition.

When materials with complex microstructure, such as foams, meet a solid wall, the standard no-slip boundary condition is not necessarily appropriate.  In particular, a no-slip boundary condition on the foam flow in a vein is unrealistic~\cite{meghdadijplmc21}.  For a foam with moderate to high liquid fractions, the same lubrication process that permits bubbles to deform and adapt to the flow of the liquid around them, and hence move past one another in the bulk, may also facilitate the motion of bubbles close to the wall of the container. It is often observed that either the layer of bubbles closest to the wall can move over any roughness at the surface (generally if they are much larger than the length-scale associated with that roughness) or for the first layer of bubbles to be trapped and the second layer to slip past the first layer~\cite{denkovsgl05,debregeastm01}. There may also be a thick wetting film at the wall, formed from interstitial liquid drawn from the foam by gravity or capillarity, which facilitates slip of the foam~\cite{kraynik88}. The surface texture can often be used to control the degree of slip~\cite{commereucRB25,rahmani}.
There is a rich variety of possible slip laws for yield stress fluids~\cite{damianou14,georgiou21,Chaparian_Tammisola_2021}. In the Navier slip law, the slip velocity at the wall is directly proportional to the stress at the wall. The constant of proportionality is related to a slip length $\beta$. Typical values of $\beta$ might be of the order of a bubble diameter; then with 20 bubbles across the channel, the slip length would be about 5\% of the channel width. Correction for wall slip in sclerotherapy relies on ex-vivo characterisation of sclerosing foams, for example in a rheometer~\cite{meghdadijplmc21}, and may not truly reflect foam behaviour in veins. It is also possible that there is a threshold wall stress which must be exceeded before a foam can slip. This is modelled by so-called stick-slip laws, which include an effective wall, or sliding, yield stress (in addition to the yield stress of the fluid itself) and employ no-slip below this threshold. One can also consider that the slip velocity (above the threshold, if one is expected to exist) may depend non-linearly on the fluid stress. In the following we use the Navier slip law, due to its simplicity for both theoretical and numerical calculations, but this may be worth revisiting in the future.

In this work, we consider two-dimensional channels with varying wall curvatures and we examine how the size and shape of the unyielded regions in a Bingham fluid depend on both the geometry and the degree of wall slip. The central unyielded plug drives the displacement flow, but additional ``dead regions'' can also form near the walls, where the fluid remains unyielded. Previous studies by Roberts and Cox~\cite{robertsc20,robertsc23} showed that curvature narrows the plug and enhances yielding, with complete yielding observed when flow transitions from straight to curved channels, and that dead zones form in highly curved sinusoidal channels, depending on the Bingham number and curvature amplitude. Here, we will demonstrate that slip at the walls further enhances fluid yielding for a given pressure gradient. Using the Bingham constitutive model and Navier slip law, and assuming a fixed pressure gradient and an effective yield stress that encapsulates foam rheology, we compute the distributions of velocity and stress in the domain. We quantify how slip modifies the unyielded plug position and extent, and assess its implications for the effectiveness of foam-based sclerotherapy.

The layout of the paper is as follows. We first describe our mathematical model for the foam flow in \S\ref{sec:maths}, including details of the assumptions on the flow, the constitutive equation, and the boundary conditions, for both straight and uniformly curved channels. The predictions of the theoretical model are described in \S\ref{sec:analysis-results}. In \S\ref{sec:numerical} we describe numerical (finite element) solutions for the flow in more complicated channel shapes, with the aim of determining the combined effect of slip and changes in wall curvature on fluid yielding. We discuss the implications of our results for foam sclerotherapy and conclude in \S\ref{sec:discussion}.

\section{Mathematical model}
\label{sec:maths}

{This section presents the mathematical model for steady, pressure-driven Bingham fluid flow in curved channels with wall slip, including the governing equations, boundary conditions, and non-dimensional formulation used to analyze velocity, stress, and yield surfaces.}

\subsection{Straight channel}
\label{sec:straight}

\begin{figure}
\centering
    \begin{subfigure}[b]{0.54\textwidth}
        \centering
        \includegraphics[width=\textwidth]{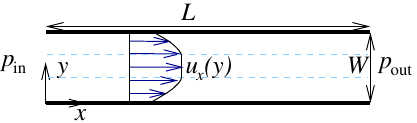}
        \caption{\;}
        \label{fig:setup_s}
    \end{subfigure}
    \begin{subfigure}[b]{0.32\textwidth}
        \centering
        \includegraphics[width=\textwidth]{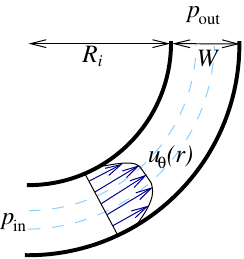}
        \caption{\;}
        \label{fig:setup_c}
    \end{subfigure}
\caption{Channel geometries. (a) Straight channel. (b) Curved channel. The yield surfaces are shown as dashed lines; the fluid is unyielded between these lines. Slip is indicated by the fluid speed being non-zero at the walls.}
\label{fig:setup}
\end{figure}

Consider first a Bingham fluid in a straight (two-dimensional) channel of length $L$ and width $W$, with inlet and outlet pressures $p_{in}$ and $p_{out}$, as illustrated in figure~\ref{fig:setup_s}. The analysis that follows  is fairly standard~\cite{birddy83,ferrasnp12,chatzimina,damianou}, although it explains the procedure that will  be used in the curved case.

\subsubsection{Governing equations}

Foam flowing at a speed of $3{\rm mm/s}$ in a vein of diameter 5mm~\cite{robertsclj21} would have a Reynolds number of order $10^{-5}$. So we assume that the Stokes equations are an appropriate description of the flow.
The fluid flow is driven by a constant pressure gradient $G=(p_{out} - p_{in})/L$. 
Stokes' equations for the slow flow of a viscous fluid state that the pressure gradient is balanced by the divergence of the stress; in this geometry, with $0 \le x \le L$ and $0\le y \le W$, the only non-zero component of the stress is $\tau_{yx}$, and so
\begin{equation}
-G = \frac{ \dd{\tau_{yx}}}{\dd{y}}.
\label{eq:stokes1}
\end{equation} 
This is coupled with the constitutive equation for the stress. For a Bingham fluid, with a yield stress $\tau_y$ and viscosity $\mu$, this is given as follows, in terms of the fluid velocity $u_x(y)$ and the appropriate component of the strain-rate, $\dd{u_{x}}/\dd{y}$:
\begin{equation}
     \begin{array}{ll} \tau_{yx} = \pm \tau_y + \mu \frac{\dd{u_{x}}}{\dd{y}}  & \mbox{ for } | \tau_{yx} | > \tau_y \\
                       \dot{\gamma}_{yx} = 0 & \mbox{ for } | \tau_{yx} | \le \tau_y. 
     \end{array}
\label{eq:constitutive1}
\end{equation} 
The sign in front of the yield stress is plus in the lower yielded region, where the stress is positive, and minus in the upper yielded region, where the stress is negative. In the centre of the channel, where the magnitude of the shear stress is below the yield stress, there is a ``plug'' of fluid with zero strain-rate.

We make the system of equations dimensionless, denoting dimensionless variables with hats, with a length-scale $W$, a velocity scale $G W^2/\mu$, and a stress scale $GW$. Then eq.~(\ref{eq:stokes1}) becomes
\begin{equation}
-1 = \frac{ \dd{\hat{\tau}_{yx}}}{\dd{\hat{y}}},
\label{eq:stokes1nondim}
\end{equation} 
and, introducing a Bingham number
\[
B = \frac{2 \tau_y}{G W},
\]
the ``yielded" part of eq.~(\ref{eq:constitutive1}), close to the walls, becomes
\begin{equation}
     \hat{\tau}_{yx} = \pm \frac{B}{2} + \frac{\dd{\hat{u}_{x}}}{\dd{\hat{y}}}  \mbox{ for } | \hat{\tau}_{yx} | > \frac{B}{2}.                   
\label{eq:constitutive1nondim}
\end{equation} 
In the no-slip case, this is subject to the boundary condition $\hat{u}_x |_{\hat{y}=0,1} = 0$. If slip is added, with slip length $\beta$, this becomes 
\begin{equation}
\hat{u}_x |_{\hat{y}=0,1} = \beta \left. \left| \hat{\tau}_{yx} \right| \right|_{\hat{y}=0,1}. 
\label{eq:slip_bc}
\end{equation}

\subsubsection{Solution}

We first integrate eq.~(\ref{eq:stokes1nondim}) to give $\hat{\tau}_{yx} = -\hat{y}+C$, where $C$ is a constant of integration. By symmetry, the stress at either wall must be the same, up to its sign; then $\left| \hat{\tau}_{yx} ({\hat{y}=0}) \right| = \left| \hat{\tau}_{yx}({\hat{y}=1}) \right| $ gives $C=1-C$, so $C=1/2$. That is, the shear stress is $\hat{\tau}_{yx} = 1/2-\hat{y}$, so it is linear across the channel.

In the region close to the lower wall, $\hat{y}=0$, eq.~(\ref{eq:constitutive1nondim}) becomes 
\begin{equation}
\frac{1}{2}-\hat{y} = \frac{B}{2}+\frac{\dd{\hat{u}_{x}}}{\dd{\hat{y}}} .
\end{equation}
Integrating gives $\hat{u}_x(\hat{y}) = \frac{1}{2} \hat{y} (1-B) - \frac{1}{2}\hat{y}^2 +C_0$, where $C_0$ is a constant of integration, to be found from the boundary condition, eq.~(\ref{eq:slip_bc}): $C_0 = \frac{1}{2}\beta$.

In the region close to the upper wall, $\hat{y}=1$, we have instead
\begin{equation}
\frac{1}{2}-\hat{y} = -\frac{B}{2}+\frac{\dd{\hat{u}_{x}}}{\dd{\hat{y}}},
\end{equation}
and so $\hat{u}_x(\hat{y}) = \frac{1}{2} \hat{y} (1+B) - \frac{1}{2}\hat{y}^2 +C_0$ here. We find $C_0 = \frac{1}{2}(\beta-B)$ from the slip boundary condition.

Equating the shear stress with the yield stress, $B/2$ in dimensionless variables, indicates that the plug occupies the region $\frac{1}{2}\left(1-B \right) \le \hat{y} \le \frac{1}{2}\left(1+B \right)$ and it therefore has width $B$ (an \emph{a posteriori} justification of our choice of the Bingham number). That is, the plug width is independent of the degree of slip at the walls. 

Hence the velocity profile in the straight channel is
\begin{equation}
  \hat{u}_x(\hat{y}) = 
   \left\{   
\begin{array}{lrcl}
  \frac{1}{2} \hat{y} (1-B) - \frac{1}{2}\hat{y}^2 + \frac{1}{2}\beta &  \mbox{ if } 0 \le & \hat{y} & < \frac{1}{2}\left(1-B \right) \\
  \frac{1}{8} (1-B)^2 + \frac{1}{2} \beta &  \mbox{ if }  \frac{1}{2}\left(1-B \right) \le &  \hat{y} & \le   \frac{1}{2}\left(1+B \right)\\
  \frac{1}{2} \hat{y} (1+B) - \frac{1}{2}\hat{y}^2 -\frac{1}{2}B + \frac{1}{2}\beta  &  \mbox{ if }   \frac{1}{2}\left(1+B \right) \le & \hat{y} & \le 1.
\end{array}
   \right.
\label{eq:velocity-straight}
\end{equation}
An example is plotted in figure~\ref{fig:velocityprofiles1}. 
The dimensionless flow-rate $\hat{Q}$ is found by integrating the velocity profile across the channel~\cite{birddy83}: 
\begin{equation}
\hat{Q} = \frac{1}{12}\left( 1- \frac{3}{2} B + \frac{1}{2}B^3 \right) + \frac{1}{2}\beta.
\label{eq:flow-rate-straight}
\end{equation}
Compared to the non-slip case, all speeds and flow-rates are increased by $\frac{1}{2}\beta$, that is, $\beta$ multiplied by the half-width of the channel. However, neither the shape of the velocity profile nor the position of the yield surfaces are affected by the slip at the walls. 
That the results are largely independent of the slip length $\beta$ is to a certain extent a consequence of the choice for the Bingham number: $B$ is set by the pressure gradient rather than by the flow-rate (or average fluid speed). Introducing slip at fixed pressure drop results in an increase in the fluid speed and hence in the shear stresses. If, instead, the Bingham number were set by the flow-rate, the pressure drop would reduce as the slip length increases, resulting in a reduction in the shear stress at the walls, and a wider plug region. Nonetheless, our choice to use the pressure gradient to define $B$ seems natural in this Poiseuille flow.

\subsection{Curved channel}
\label{sec:curved-deriv}

We now consider a uniformly \emph{curved} channel, of width $W$, as illustrated in figure~\ref{fig:setup_c}, and work in polar coordinates $(r,\theta)$. The curvature of the channel is defined to be $\kappa = 1/R_i$, where $R_i$ denotes the radius of the inner wall of the channel. As $\kappa \rightarrow 0$ (i.e. the inner radius becomes very large) we recover the results for a straight channel. We write $R_o = R_i+W$ for the radius of the outer wall, $R_c = R_i+W/2$ for the position of the centreline, and $\theta_c$ for the total turning angle of the channel. The derivation closely follows the method above, but the results have a non-trivial dependence on the slip length $\beta$.

\subsubsection{Governing equations}

In this case the pressure gradient driving the flow is $G=(p_{out} - p_{in})/(R_c \theta_c)$. In the limit of small Dean number that we consider here there is again only one non-zero component of the stress, $\tau_{r\theta}$, and so, following eq.~(\ref{eq:stokes1}), we have
\begin{equation}
-\frac{R_c G}{r} = \frac{1}{r^2} \frac{ \dd{\;} }{\dd{r}}\left( r^2 \tau_{r\theta} \right).
\label{eq:stokes1r}
\end{equation} 
Similarly, the constitutive equation in the yielded parts of the flow is
\begin{equation}
   \tau_{r\theta} = \pm \tau_y + \mu r \frac{\dd{\;}}{\dd{r}} \left(\frac{u_\theta}{r} \right)  \mbox{ for } | \tau_{r\theta} | > \tau_y,
\label{eq:constitutive1r}
\end{equation} 
where there is only one component of the fluid speed, $u_\theta(r)$.

We take the same non-dimensionalisation as for the straight-channel case, with $\hat{\kappa} = W/R_i$. Then eq.~(\ref{eq:stokes1r}) for the stress becomes
\begin{equation}
-\left(\frac{1}{\hat{\kappa}} + \frac{1}{2}\right) = \frac{1}{\hat{r}} \frac{ \dd{\;} }{\dd{\hat{r}}}\left( \hat{r}^2 \hat{\tau}_{r\theta} \right),
\label{eq:stokes1rnondim}
\end{equation} 
and the constitutive equation, eq.~(\ref{eq:constitutive1r}), becomes
\begin{equation}
   \hat{\tau}_{r\theta} = \pm \frac{B}{2} + \hat{r} \frac{\dd{\;}}{\dd{r}} \left(\frac{\hat{u}_\theta}{\hat{r}} \right)  \mbox{ for } | \hat{\tau}_{r\theta} | > \frac{B}{2}.
\label{eq:constitutive1rnondim}
\end{equation}
As before, the strain-rate is zero in the unyielded plug near the middle of the channel.
The boundary conditions on the walls of the channel are 
\begin{equation}
\hat{u}_\theta |_{\hat{r}=1/\hat{\kappa},1+1/\hat{\kappa}} = \beta \left. \left| \hat{\tau}_{r\theta} \right| \right|_{\hat{r}=1/\hat{\kappa},1+1/\hat{\kappa}},
\label{eq:slip_bcr}
\end{equation}
which is clearly less straightforward to evaluate than the straight-channel case since we don't yet have a simple expression for the stress.

\subsubsection{Solution}

We first solve eq.~(\ref{eq:stokes1rnondim}) for the stress:
\begin{equation}
\hat{\tau}_{r\theta} = -\left(\frac{2+\hat{\kappa} }{4 \hat{\kappa}}\right)+\frac{C}{\hat{r}^2},
\end{equation}
where $C$ is a constant of integration.
There are again two yield surfaces, an inner one (at $\hat{r} = \hat{r}_i$) and an outer one (at $\hat{r} = \hat{r}_o$). These are found where $|\hat{\tau}_{r\theta}| = \frac{1}{2}B$:
\begin{equation}
\hat{r}_i^2 = \frac{4 C \hat{\kappa}}{2+\hat{\kappa}+2 \hat{\kappa}B} 
\mbox{ and }
\hat{r}_o^2 = \frac{4 C \hat{\kappa}}{2+\hat{\kappa} - 2 \hat{\kappa}B}.
\label{eq:r_i_r_o_defn}
\end{equation}
We choose to use these equations to eliminate $C$, and write the stress in terms of the positions of the inner and outer yield surfaces. These will later be found through matching the solution for the velocity at these points. We define the augmented Bingham numbers 
\[
 B^{\pm} = \frac{1}{2} \left( \frac{2+\hat{\kappa} }{2 \hat{\kappa}} \pm B \right),
\]
and then in the inner region the stress is
\begin{equation}
\hat{\tau}_{r\theta} = -\left(\frac{2+\hat{\kappa} }{4 \hat{\kappa}}\right)+B^{+}\frac{\hat{r}_i^2}{\hat{r}^2},
\end{equation}
and in the outer region it is
\begin{equation}
\hat{\tau}_{r\theta} = -\left(\frac{2+\hat{\kappa} }{4 \hat{\kappa}}\right)+B^{-}\frac{\hat{r}_o^2}{\hat{r}^2}.
\end{equation}
The stress is therefore highest at the inner wall and decays as $1/\hat{r}^2$ across the channel. Between the inner and outer regions the strain-rate is zero, and the fluid moves in solid body rotation there, with a velocity profile of the form $\hat{u}_\theta = A \hat{r}$ for some constant $A$.

We now find the fluid speed in each of the inner and outer regions by substituting for the stress in eq.~(\ref{eq:constitutive1rnondim}) and integrating. In the inner region we have
\begin{equation}
\hat{u}_\theta  = -B^+ \hat{r} \ln(\hat{r}) - \frac{1}{2} B^+ \frac{\hat{r}_i^2}{\hat{r}} + C_i \hat{r},
\label{eq:velocity_curved_inner}
\end{equation}
where $C_i$ is a constant of integration. Similarly, in the outer region
\begin{equation}
\hat{u}_\theta  = -B^- \hat{r} \ln(\hat{r}) - \frac{1}{2} B^- \frac{\hat{r}_o^2}{\hat{r}} + C_o \hat{r},
\label{eq:velocity_curved_outer}
\end{equation}
where $C_o$ denotes the constant of integration. These constants are found from the slip boundary condition, eq.~(\ref{eq:slip_bcr}):
\begin{eqnarray}
C_i & = & 
  \frac{1}{2} B^+ \hat{r}_i^2 \hat{\kappa}^2 \left( 1 + 2 \beta \hat{\kappa} \right)  
  + B^+ \ln\left(\frac{1}{\hat{\kappa}}\right)
  -\frac{1}{4}\beta \left(\hat{\kappa}+2\right), 
\\
C_o & = & 
   \frac{1}{2} B^- \hat{r}_o^2 \left(\frac{\hat{\kappa}}{\hat{\kappa}+1}\right)^2 \left( 1 + 2 \beta \left(\frac{\hat{\kappa}}{\hat{\kappa}+1}\right) \right)  
  + B^- \ln\left(\frac{1}{\hat{\kappa}}+1\right)
  -\frac{1}{4}\beta \frac{\left(\hat{\kappa}+2\right)}{\left(\hat{\kappa}+1\right)}.
\end{eqnarray}
The constant $A$ is found by matching the fluid speed at (for example) the inner yield surface:
\begin{equation}
A = \frac{1}{2} B^+ \left(\hat{\kappa}^2 \hat{r}_i^2 - 1\right) 
- B^+ \ln(\hat{\kappa} \hat{r}_i ) +
\beta \hat{\kappa} \left(  B^+ \hat{\kappa}^2 \hat{r}_i^2  - \frac{\hat{\kappa}+2}{4\hat{\kappa}}\right) .
\end{equation}
We write $\hat{\kappa}^\prime = \displaystyle \frac{\hat{\kappa}}{\hat{\kappa}+1}$, and then the velocity profile is
\begin{equation}
  \hat{u}_\theta({\hat{r}}) = 
   \left\{   
\begin{array}{lrcl}
 \frac{1}{2} B^+ \hat{r}_i^2 \left(\hat{\kappa}^2 \hat{r} - \displaystyle\frac{1}{\hat{r}} \right)
  -B^+ \hat{r} \ln({\hat{\kappa}\hat{r}})
 + \beta \hat{r} \hat{\kappa}\left( B^+ \hat{r}_i^2 \hat{\kappa}^2 -\displaystyle \frac{\hat{\kappa}+2}{4\hat{\kappa}}  \right)  
  &  \mbox{ if } \frac{1}{\hat{\kappa}} \le & \hat{r} & < \hat{r}_i \\
  
  \frac{1}{2} B^+\hat{r}  \left(\hat{\kappa}^2 \hat{r}_i^2 - 1\right) - B^+  \hat{r} \ln( \hat{\kappa}\hat{r}_i) + \beta \hat{r}  \hat{\kappa} \left(  B^+ \hat{r}_i^2 \hat{\kappa}^2 - \displaystyle \frac{\hat{\kappa}+2}{4\hat{\kappa}} \right)  &  \mbox{ if }  \hat{r}_i \le &  \hat{r} & \le  \hat{r}_o \\
  \frac{1}{2} B^-\hat{r}_o^2 \left(\hat{\kappa}^{\prime^2} \hat{r} - \displaystyle \frac{1}{\hat{r}}  \right)
  -B^- \hat{r} \ln\left( \hat{\kappa}^\prime \hat{r}  \right)
  +\beta  \hat{r} \hat{\kappa}^\prime
  \left( B^- \hat{r}_o^2 \hat{\kappa}^{\prime^2} - \displaystyle \frac{\hat{\kappa}+2}{4\hat{\kappa}}\right)  
  &  \mbox{ if }   \hat{r}_o \le & \hat{r} & \le \frac{1}{\hat{\kappa}^\prime}.
\end{array}
   \right.
\label{eq:velocity_curved}
\end{equation}
Note that the slip term acts as a solid-body rotation across the three parts of the solution.

Finally, we find the positions of the yield surface positions $\hat{r}_i$ and $\hat{r}_o$, which depend on the slip length $\beta$. From eq.~(\ref{eq:r_i_r_o_defn}) we have $B^- \hat{r}_o^2 = B^+ \hat{r}_i^2$, 
and then there is a further matching condition at the outer yield surface to give a second relationship between $\hat{r}_i$ and $\hat{r}_o$:
\begin{eqnarray}
\frac{1}{2} B^- \left(   \hat{\kappa}^{\prime^2} \hat{r}_o^2 -1\right) 
- B^- \ln \left( \hat{\kappa}^\prime \hat{r}_o  \right)
+  \beta \hat{\kappa}^\prime \left( B^- \hat{\kappa}^{\prime^2}  \hat{r}_o^2  - \frac{\hat{\kappa}+2}{4\hat{\kappa}}\right) 
= \nonumber \\
\frac{1}{2} B^+ \left( \hat{\kappa}^2 \hat{r}_i^2  -1\right) 
- B^+ \ln \left( \hat{\kappa} \hat{r}_i  \right)
+  \beta \hat{\kappa}   
\left( 
 B^+ \hat{\kappa}^2 \hat{r}_i^2  
 -  \displaystyle \frac{\hat{\kappa}+2}{4\hat{\kappa}}
\right) 
.
\label{eq:final-condition}
\end{eqnarray}
As in the no-slip case~\cite{robertsc20}, the position of either of the yield surfaces can be expressed in the form of a Lambert W function~\cite{corless96,pitsiliou}. We choose instead to solve eq.~(\ref{eq:final-condition}) numerically to compute the velocity profile, which is straightforward with a python (scipy) root-finding routine~\cite{scipy}, and so the results below are referred to as ``semi-analytic".

\subsection{Semi-analytical results}
\label{sec:analysis-results}

\begin{figure}
\centering
    \begin{subfigure}[c]{0.48\textwidth}
        \centering
	\includegraphics[width=\textwidth]{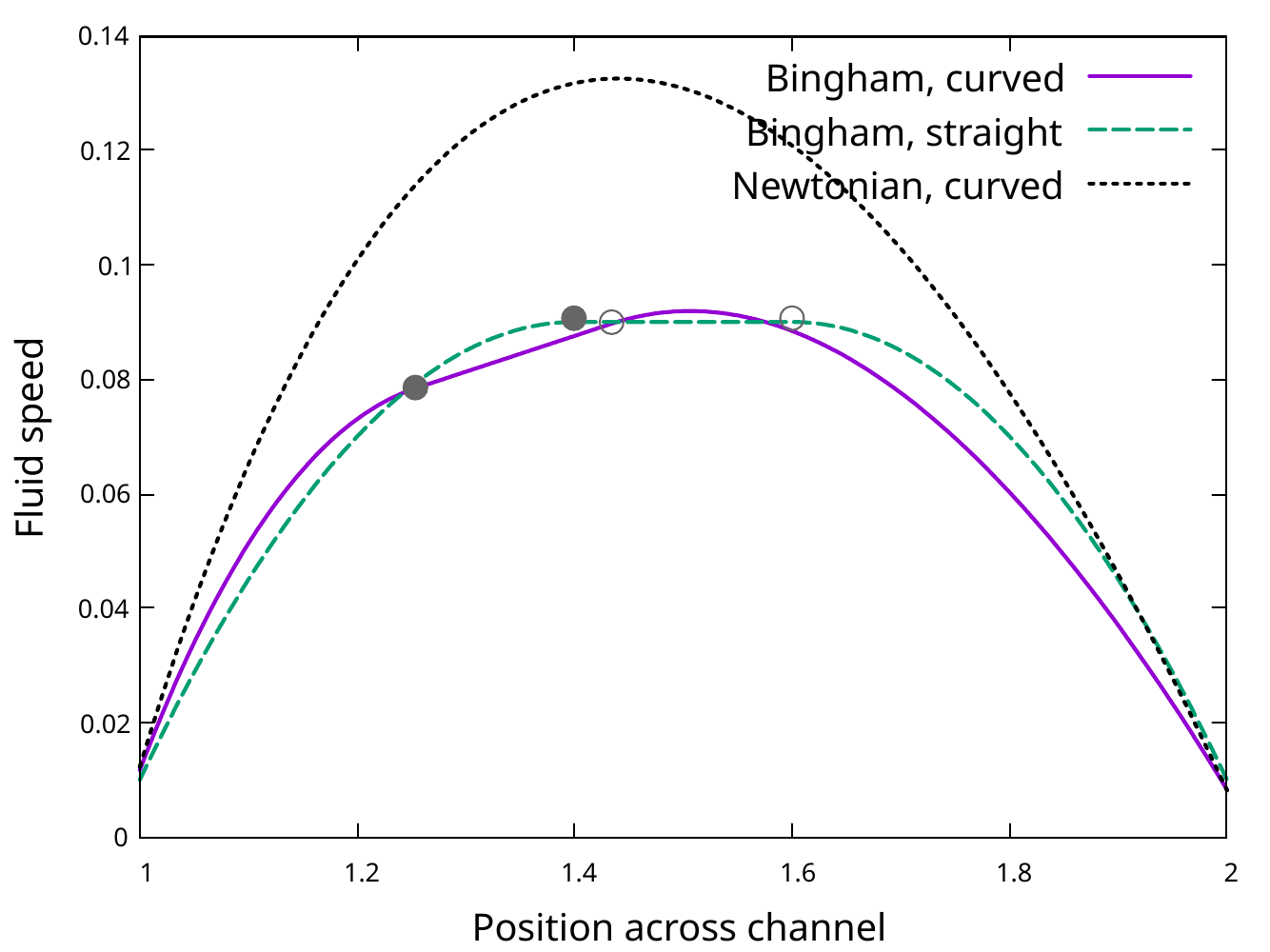}
        \caption{\;}
        \label{fig:velocity_profiles_1a}
    \end{subfigure}
    \begin{subfigure}[c]{0.48\textwidth}
        \centering
	\includegraphics[width=\textwidth]{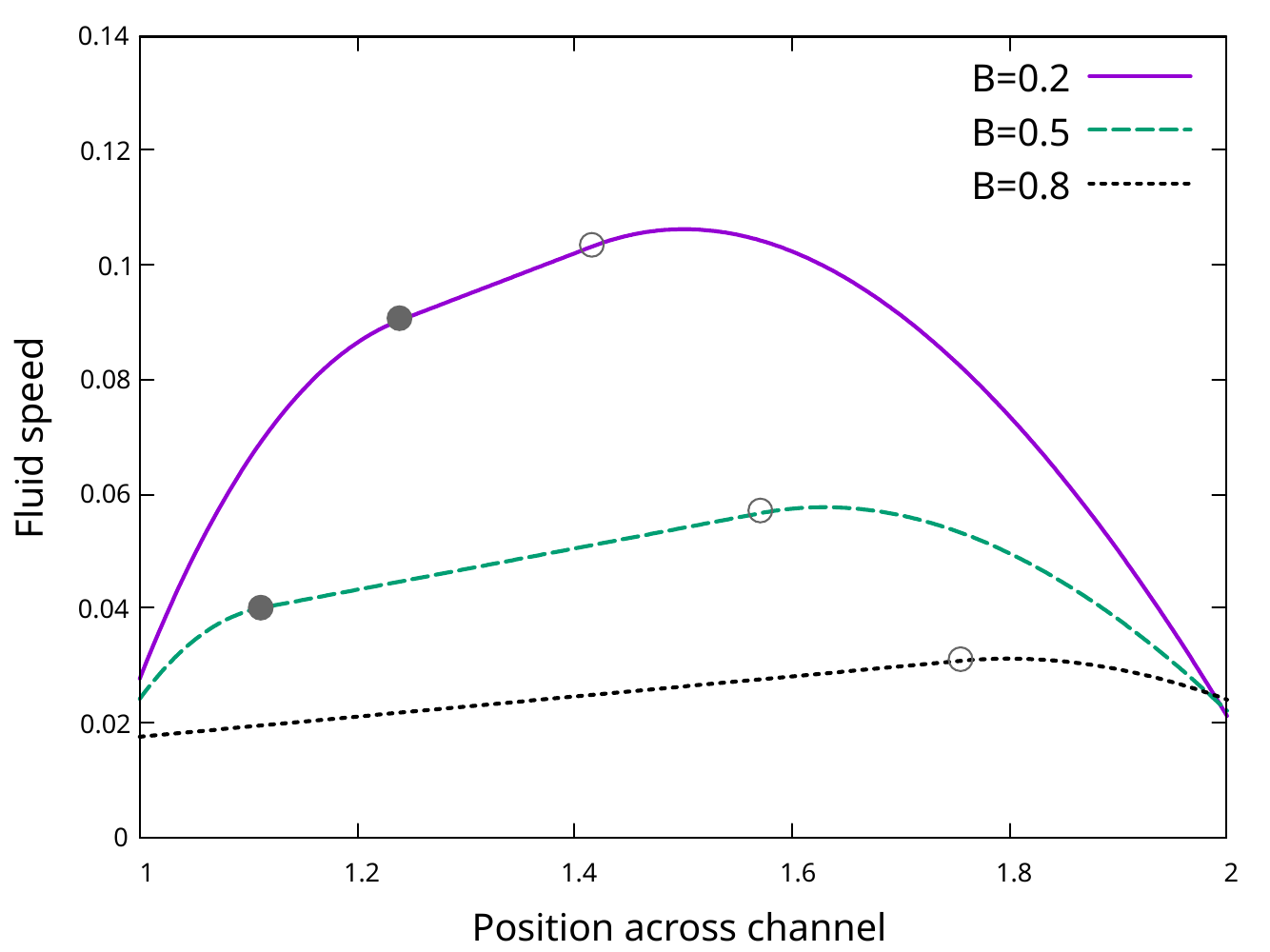}
        \caption{\;}
        \label{fig:velocity_profiles_1b}
    \end{subfigure}

    \begin{subfigure}[c]{0.48\textwidth}
        \centering
	\includegraphics[width=\textwidth]{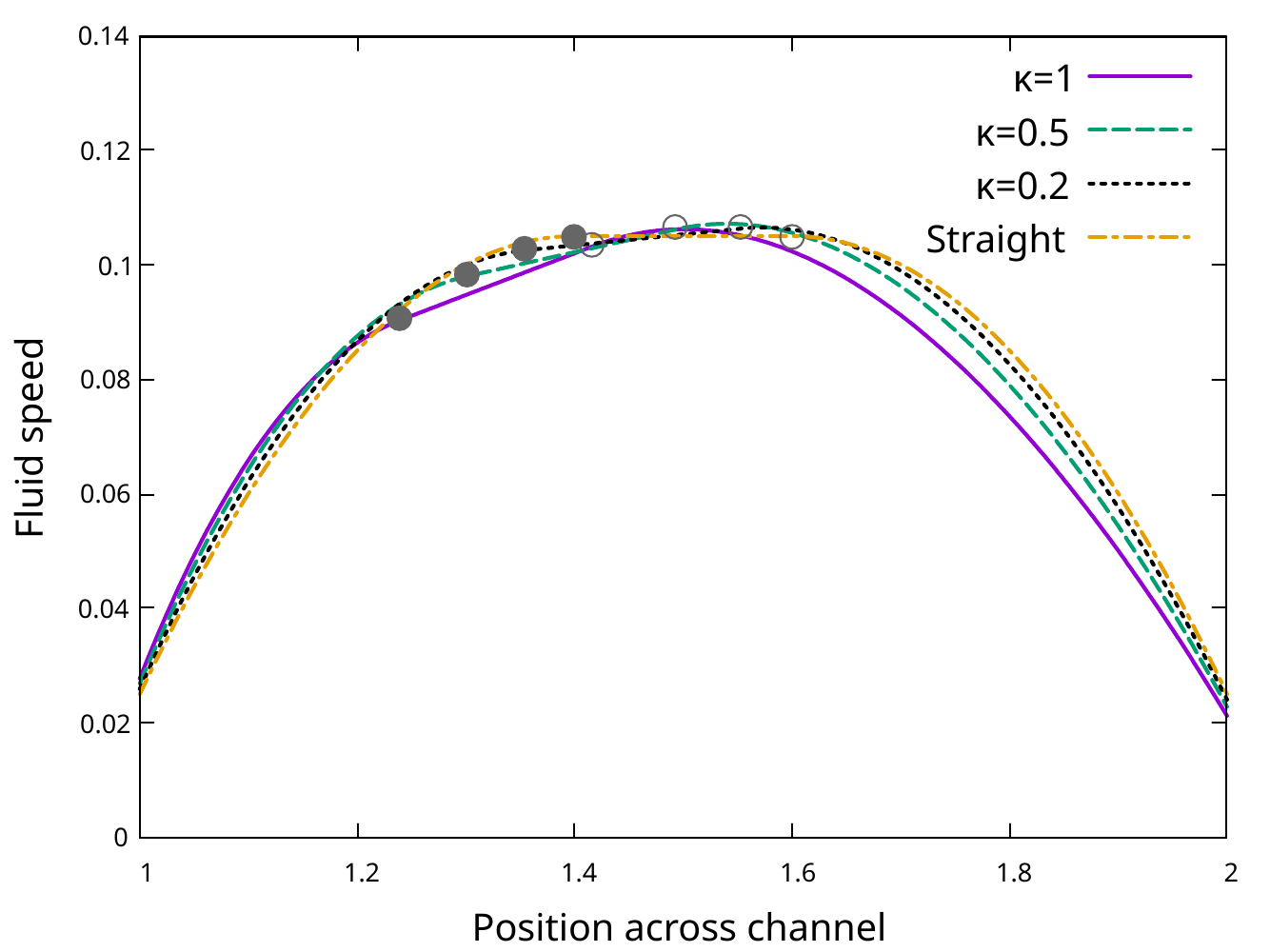}
        \caption{\;}
        \label{fig:velocity_profiles_1c}
    \end{subfigure}
    \begin{subfigure}[c]{0.48\textwidth}
        \centering
	\includegraphics[width=\textwidth]{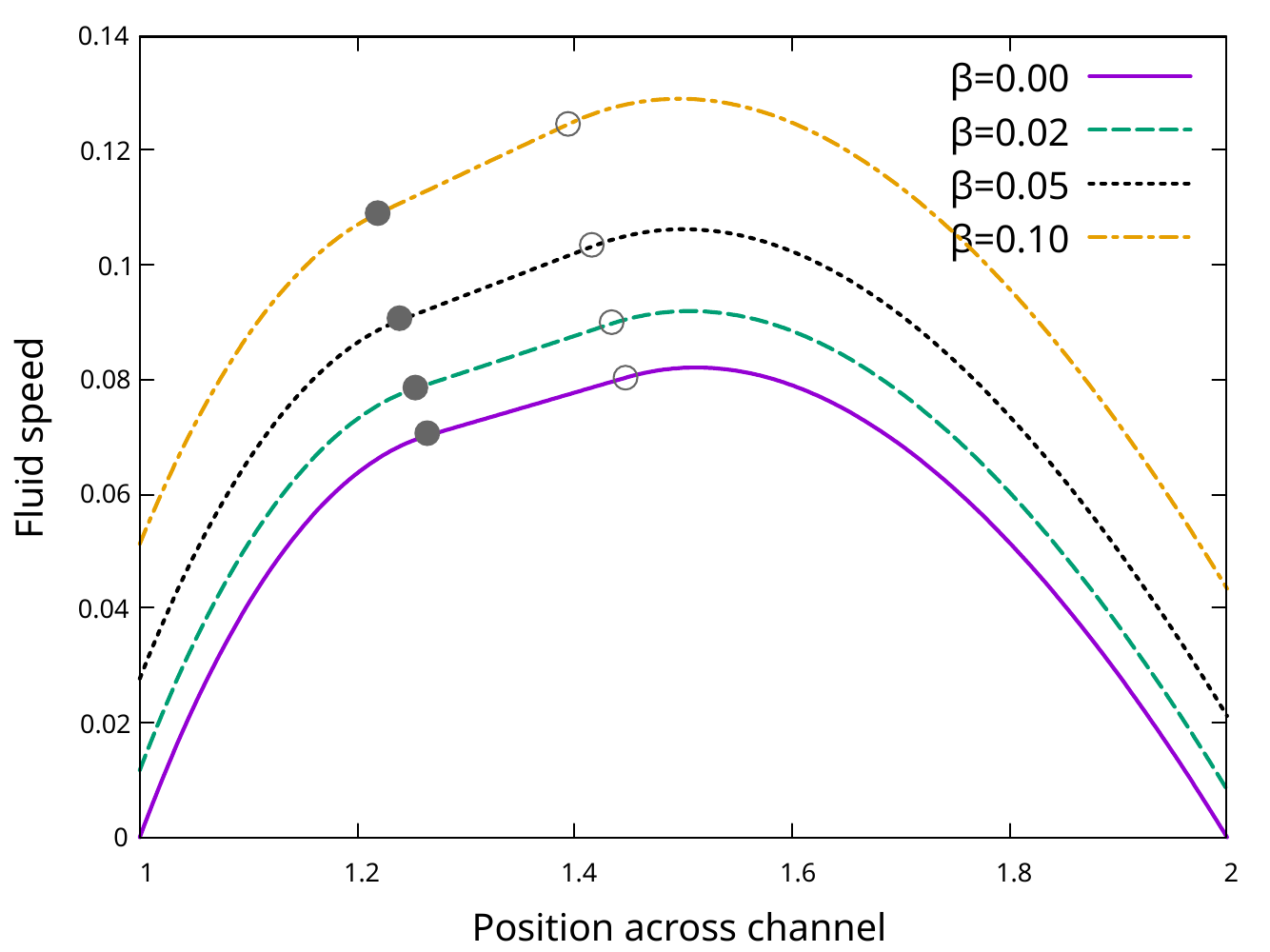}
        \caption{\;}
        \label{fig:velocity_profiles_1d}
    \end{subfigure}
\caption{
Velocity profiles for pressure-driven flow of a Bingham fluid. 
Small dots marks the positions of the yield surfaces.
(a) Profile in a curved channel with unit inner curvature ($\hat{\kappa}=1$), from eq.~(\ref{eq:velocity_curved}).
The slip parameter is $\beta = 0.02$ and the Bingham number is $B=0.2$.
Also shown for comparison are the velocity profile of a Newtonian fluid ($B=0$) in a curved channel, and of a Bingham fluid in a straight-channel (which is shifted from $0\le \hat{y} \le 1$ to $1\le \hat{r} \le 2$) from eq.~(\ref{eq:velocity-straight}). 
(b) Variation with Bingham number. Here $\hat{\kappa}=1$ and $\beta = 0.05$.
(c) Variation with channel curvature $\hat{\kappa}$. Here $\beta=0.05$ and $B=0.2$. All positions are shifted to the interval [0,1] to enable direct comparison.
(d) Variation with slip length $\beta$. Here $\hat{\kappa}=1$ and $B=0.2$.
}
\label{fig:velocityprofiles1}
\end{figure}

\begin{figure}
\centering
    \begin{subfigure}[c]{0.48\textwidth}
        \centering
	\includegraphics[width=\textwidth]{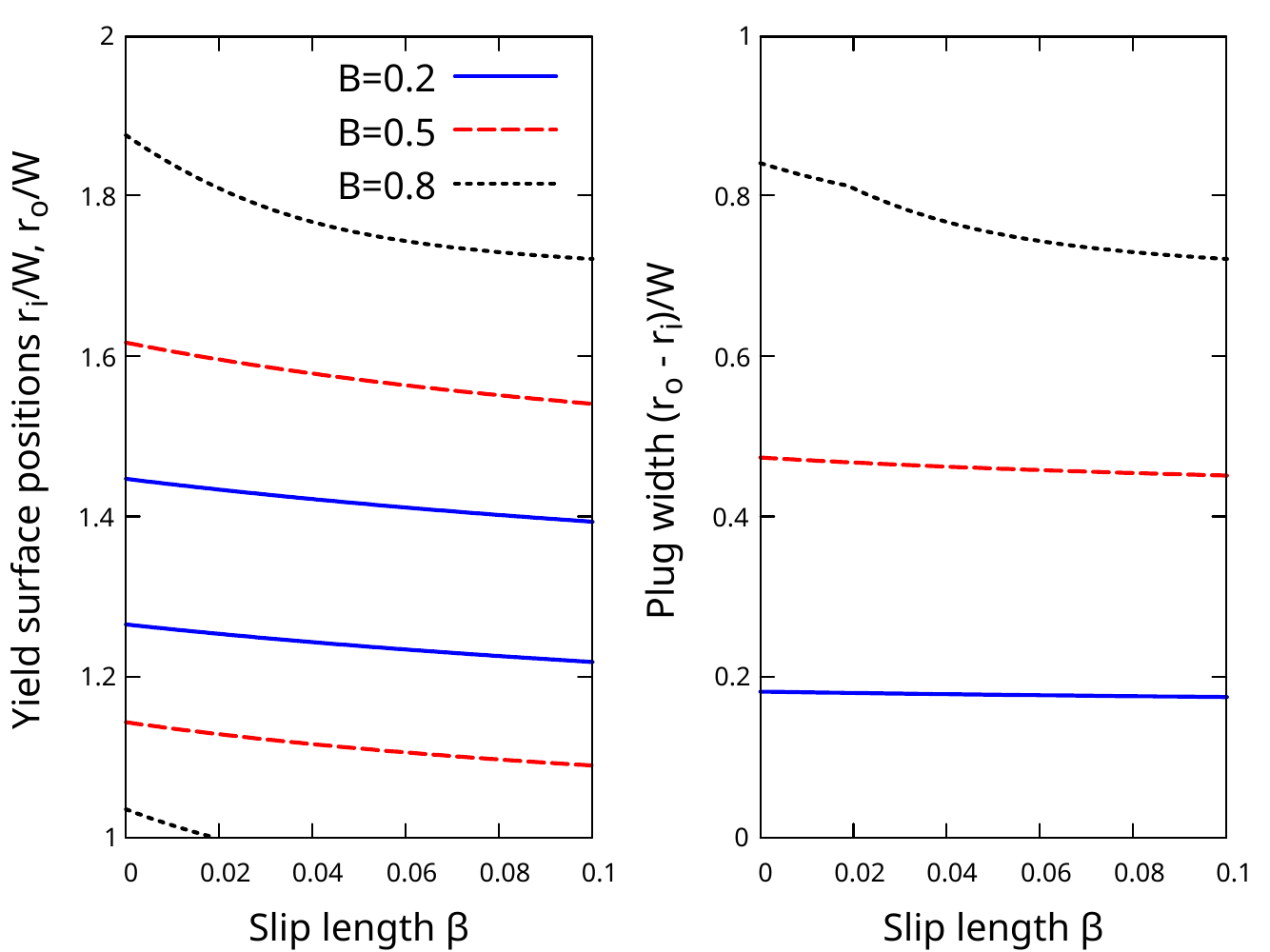}
        \caption{\;}
        \label{fig:sliplength1a}
    \end{subfigure}
    \begin{subfigure}[c]{0.48\textwidth}
        \centering
	\includegraphics[width=\textwidth]{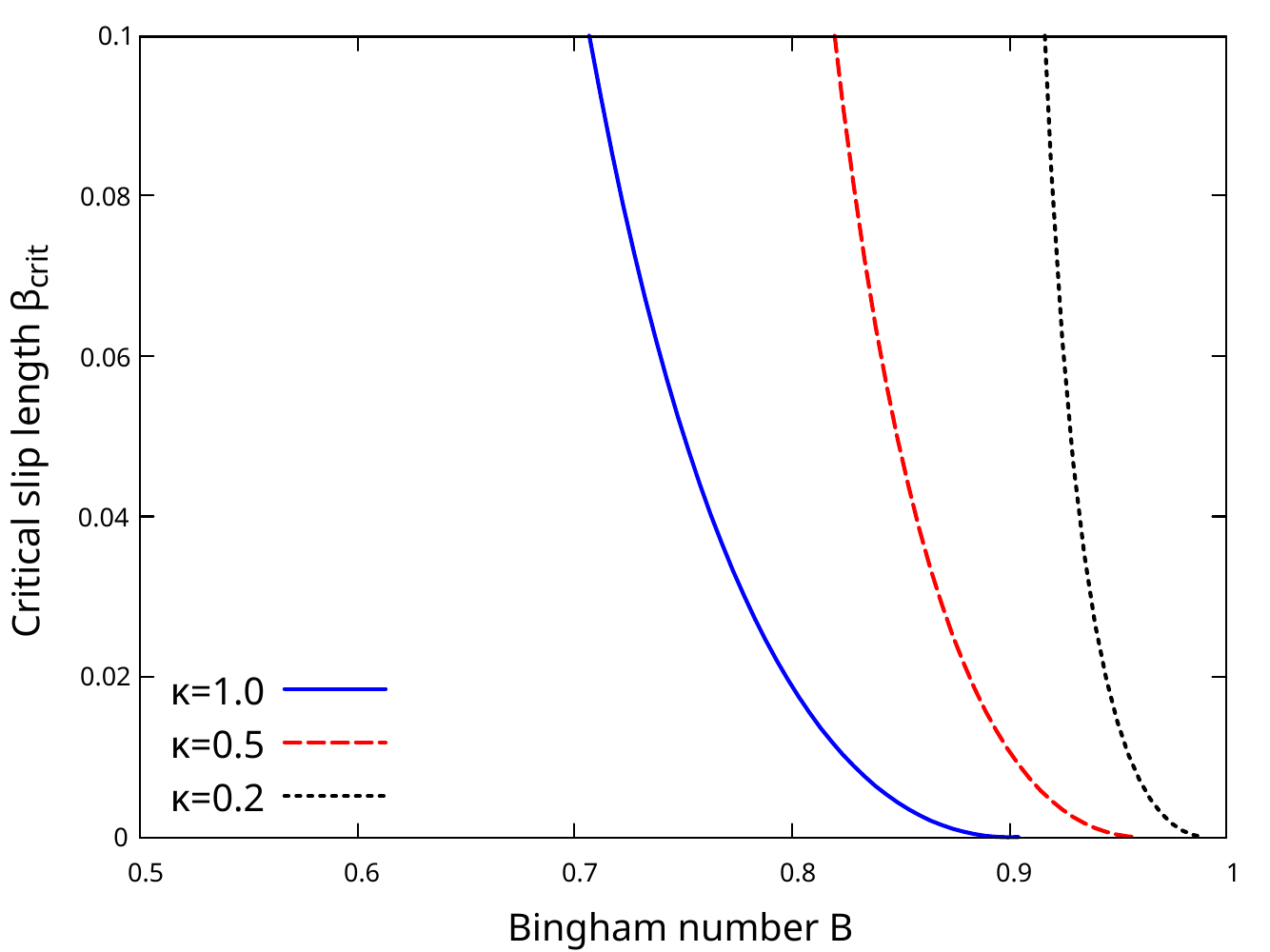}
        \caption{\;}
        \label{fig:sliplength1b}
    \end{subfigure}
\caption{(a) The effect of varying the slip length, with $\hat{\kappa}=1$, on the position of the yield surfaces and the plug width. 
Recall that for a straight channel the plug width is equal to $B$.
(b) The critical slip length $\beta_{\rm crit}$, from eq.~(\ref{eq:beta_crit}), for three different channel curvatures.
}
\label{fig:sliplength1}
\end{figure}

An example of these velocity profiles is shown in figure~\ref{fig:velocity_profiles_1a}, which compares the straight and curved channel solutions for a Bingham fluid. We also compare with the Newtonian case, which has $B=0$, so $B^+ = B^- = (\hat{\kappa}+2)/(4\hat{\kappa})$ and $\hat{r}_i = \hat{r}_o$. The effect of channel curvature is to ``tilt" the straight section of the profile representing the unyielded plug, and to reduce its width slightly. As the channel curvature decreases (figure~\ref{fig:velocity_profiles_1c}) the tilt of the flat part of the profile decreases, and the curved-channel solutions reduce to the straight ones.
Figure~\ref{fig:velocity_profiles_1b} shows that as the Bingham number increases, the fluid speed decreases and the plug widens. The fluid speed is slightly greater at the inner wall, and decreases slightly with increasing $B$. Around $B=0.8$ the inner yield surface meets the inner wall, and only the fluid in the outer part of the channel is sheared. 
Figure~\ref{fig:velocity_profiles_1d} shows the change in velocity profile as the slip length changes. Recall that the pressure gradient is fixed here, since the Bingham number is fixed. In addition to the expected increase in the magnitude of the velocity with increasing slip length, the shape of the velocity profile also changes slightly.  The changes to the width of the plug are shown more clearly in figure~\ref{fig:sliplength1a}: the yield surfaces move slightly towards the inner wall and the plug width decreases slightly as $\beta$ increases.  These differences are reduced when the radius of curvature of the channel $R_i$ increases, that is, as the channel curvature is reduced and the results approach those for the straight channel.

For $B=0.8$ and $\hat{\kappa}=1$ the inner yield surface reaches the inner wall $\hat{r} = 1$ ($r=R_i$) when the slip length has increases to $\beta \approx 0.02$ (figure~\ref{fig:sliplength1a}). The fluid continues to slip at the inner wall, but moves in solid body rotation all the way out to the outer yield surface, at a speed determined by the outer flow. For smaller $B$ (and smaller curvatures) this response is less likely, although we should expect that the critical value, $\beta_{\rm crit}(B,\hat{\kappa})$, for the inner yield surface to reach the inner wall should decrease as $B$ increases beyond some threshold value. In general, we must solve eq.~(\ref{eq:final-condition}) with $\hat{r}_i = 1/{\hat{\kappa}}$ and $\hat{r}_o = \sqrt{B^+/B^-}/\hat{\kappa}$. We find
\begin{equation}
\beta_{\rm crit} =  \frac{		 
 \displaystyle \frac{1}{(\hat{\kappa}+1)^2} B^+ -B^- -B^- \ln\left(\displaystyle \frac{B^+}{B^-}\frac{1}{(\hat{\kappa}+1)^2} \right)	}  
  {    
  2 \hat{\kappa} B^+ \left( 1+ \displaystyle \frac{1}{(\hat{\kappa}+1)^3} \right) - \displaystyle \frac{(\hat{\kappa}+2)^2}{2(\hat{\kappa}+1)}       
  }.
\label{eq:beta_crit}
\end{equation}
This is shown in figure~\ref{fig:sliplength1b} for three values of the wall curvature. As the Bingham number increases the critical slip length decreases to zero. 
For a given slip length, the value of the Bingham number at which the inner yield surface meets the inner wall decreases as the curvature increases.

For small Bingham number, the inner yield surface never reaches the inner wall, however large the slip length. That is, the denominator of the expression for $\beta_{\rm crit}$ in eq.~(\ref{eq:beta_crit}) can be zero for small $B$. We find that the minimum value of the Bingham number for which this can happen is $B_{\rm min} = (\hat{\kappa}+2)^2/(2(1+(\hat{\kappa}+1)^3))$; this is equal to $\frac{1}{2}$ for $\hat{\kappa} = 1$ and tends to one as $\hat{\kappa}$ tends to zero. That is, in the case of a straight channel, the yield surfaces only reach the walls for $B=1$.

Recall that in a straight channel the plug width is equal to the value of the Bingham number $B$ (for our choice of Bingham number dependent on pressure gradient). In a curved channel, we expect the plug width to be less than $B$, because the curvature induces greater yielding, which is seen to be the case in figure~\ref{fig:sliplength1a}. For $B=0.8$ however, the largest Bingham number shown, the plug width is greater than $B$ for small slip lengths. This is an indication that for slightly large $B$ (still less than one) the flow will cease, because the plug spans the whole channel. By construction, the numerator of eq.~(\ref{eq:beta_crit}) is zero when the outer yield surface meets the outer wall ($r_o = R_o$), and so the condition for the flow stopping is equivalent to finding the Bingham number for which $\beta_{\rm crit} = 0$. More straightforwardly, this occurs when $R_o^2 B^- = R_i^2 B^+$. This gives the critical Bingham number for flow in the channel to stop as
\begin{equation}
B_{\rm crit} = \frac{\left( \hat{\kappa} +2\right)^2}{2\left(2+2\hat{\kappa}+\hat{\kappa}^2\right)}.
\end{equation} 
This clearly depends on the curvature: it is equal to 0.9 for $\hat{\kappa} = 1$, as is evident from figure~\ref{fig:sliplength1b}, and increases quickly towards one as the channel curvature reduces.

\section{Changes in channel curvature}
\label{sec:numerical}

When a valve in a vein ceases to function correctly, blood collects and causes the vein to enlarge and deform. This can induce significant curvature in the vein, and thus it is of interest to determine the extent of the plug region in a curved vein. 
We now examine the effect of changes in channel curvature on the yielding of the fluid, combined with the effects of slip, using a numerical procedure. 

Pressure-driven flows of yield stress fluids in non-uniformly curved channels have received relatively little attention. Closed-form expressions for the flows are difficult to find, hindering the validation of numerical solutions.
In two dimensions, in the no-slip case, Roustaei and Frigaard~\cite{roustaei2013occurrence} used the augmented Lagrangian method to find the plug regions in flow through a ``wavy-walled" channel, and Roberts and Cox studied the flow in both an abrupt transition from a straight to a curved channel~\cite{robertsc20} and a sinusoidal channel~\cite{robertsc23} using regularisation methods. In three dimensions, again for no-slip flows, Lattice-Boltzmann~\cite{limuxiong} and regularisation methods~\cite{sutton} have been used for Herschel-Bulkley fluids in $90^\circ$ bends in pipes with circular cross-section.

We describe our finite-element calculations using regularisation in \S\ref{sec:numerical-method}. In \S\ref{sec:straight-to-curved} we determine the flow in the transition regions between a straight and a curved channel, for which we know the limiting behaviour far from the ``join" and can therefore validate our numerical solutions. In this geometry there is just one, abrupt, change in wall curvature. Therefore in \S\ref{sec:sinusoidal} we consider a sinusoidal-shaped channel in which the curvature not only varies along the channel but can also change sign. We will show that  any dead regions that are found in the no-slip case rapidly shrink and disappear as the slip length increases.
 
\subsection{Numerical method}
\label{sec:numerical-method}

We use the finite element software FreeFem++~\cite{freefem},  extending the method described in~\cite{robertsc23}, which we summarise briefly here, to include the Navier  slip boundary condition. The calculations are performed in dimensionless units, with (in effect) channel width $W$, fluid viscosity $\mu$ and pressure gradient $G$ all equal to one. The software solves the equations of motion of a viscous fluid using the weak formulation, with a regularisation of the stress (cf eq.~(\ref{eq:constitutive1})) due to Papanastasiou ~\cite{papanastasiou87}. The criterion on the stress~\cite{frigaardn05} for the yield surface positions is dependent on the Bingham number and the smoothing of the regularisation. The channel geometry is meshed with about $10^5$ triangles, and the equations are solved with P3 elements for velocity and P1 for pressure.

The slip boundary condition is imposed in two parts: by penalising flow normal to the boundary and by solving an integral condition for the tangential component of the velocity (which includes the slip length $\beta$)~\cite{Lund2015}.

\subsection{Straight-to-curved channel}
\label{sec:straight-to-curved}

The first geometry that we consider is a straight channel joined on to a curved channel, for which we can validate the numerical method against the results in eqs~(\ref{eq:velocity-straight}) and (\ref{eq:velocity_curved}) away from the join. The straight channel has length $L$ equal to twice its width ($L=2$) and meets one quarter of an annular channel ($\theta_c = \pi/2$), effectively joining the two diagrams of figure~\ref{fig:setup}.  Hence there is an abrupt change in curvature, from zero to $\kappa=1/R_i$. 

\begin{figure}
\centering
    \begin{subfigure}[c]{0.3\textwidth}
        \centering
        \includegraphics[width=\textwidth]{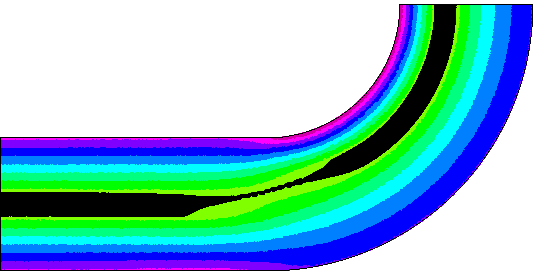}
        \caption{\;}
        \label{fig:comp-yield1a}
    \end{subfigure}
    \begin{subfigure}[c]{0.3\textwidth}
        \centering
        \includegraphics[width=\textwidth]{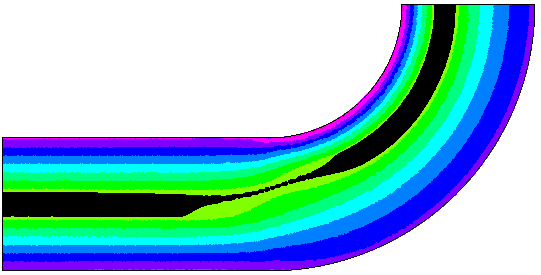}
        \caption{\;}
        \label{fig:comp-yield1b}
    \end{subfigure}
    \begin{subfigure}[c]{0.31\textwidth}
        \centering
        \includegraphics[width=\textwidth]{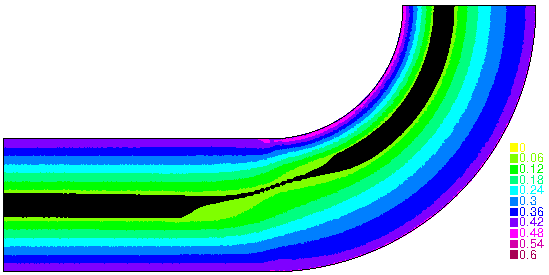}
        \caption{\;}
        \label{fig:comp-yield1c}
    \end{subfigure}

\centering
    \begin{subfigure}[c]{0.3\textwidth}
        \centering
        \includegraphics[width=\textwidth]{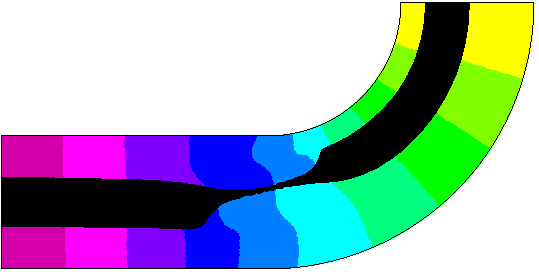}
        \caption{\;}
        \label{fig:comp-yield2a}
    \end{subfigure}
    \begin{subfigure}[c]{0.3\textwidth}
        \centering
        \includegraphics[width=\textwidth]{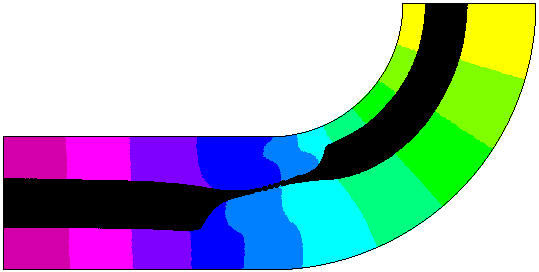}
        \caption{\;}
        \label{fig:comp-yield2b}
    \end{subfigure}
    \begin{subfigure}[c]{0.31\textwidth}
        \centering
        \includegraphics[width=\textwidth]{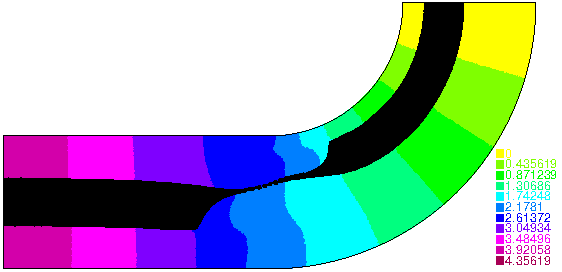}
        \caption{\;}
        \label{fig:comp-yield2c}
    \end{subfigure}

\centering
    \begin{subfigure}[c]{0.3\textwidth}
        \centering
        \includegraphics[width=\textwidth]{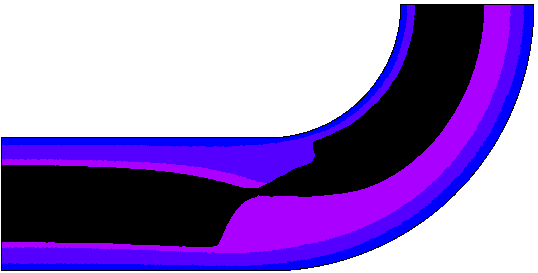}
        \caption{\;}
        \label{fig:comp-yield3a}
    \end{subfigure}
    \begin{subfigure}[c]{0.3\textwidth}
        \centering
        \includegraphics[width=\textwidth]{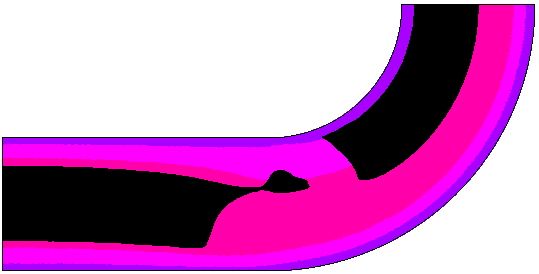}
        \caption{\;}
        \label{fig:comp-yield3b}
    \end{subfigure}
    \begin{subfigure}[c]{0.31\textwidth}
        \centering
        \includegraphics[width=\textwidth]{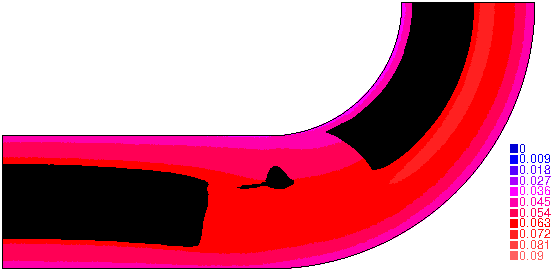}
        \caption{\;}
        \label{fig:comp-yield3c}
    \end{subfigure}
\caption{Numerical solution for a pressure-driven Bingham fluid in a channel that changes from straight to curved with curvature $\kappa = 1$.  (a)-(c) Bingham number $B=0.2$, slip length $\beta = 0.02$, 0.06 and 0.1 respectively. The contours shown outside the black unyielded plug are for the magnitude of the stress.  (d)-(f) Same images for $B=0.4$, with contours of pressure. (g)-(i) Same  images for $B = 0.6$, with contours of speed.  }
\label{fig:comp-yield1}
\end{figure}

Examples of the predicted shape of the plug region are shown in figure~\ref{fig:comp-yield1}. At the entrance and exit of the channel we recover the velocity profile  and plug width in the straight and curved channel derived above (data not shown). In every case there is almost complete yielding of the fluid in the transition region between the two parts of the channel, as was found in the no-slip case~\cite{robertsc20}. The shape of the plug where it ends and re-starts is repeated in each case: it tapers asymmetrically, with the width remaining constant but the plug moving towards the outer wall, in the straight part of the channel; in the curved part of the channel the opposite occurs, with a shift towards the inner wall.

\begin{figure}
\centering
    \begin{subfigure}[c]{0.48\textwidth}
        \centering
        \includegraphics[width=\textwidth]{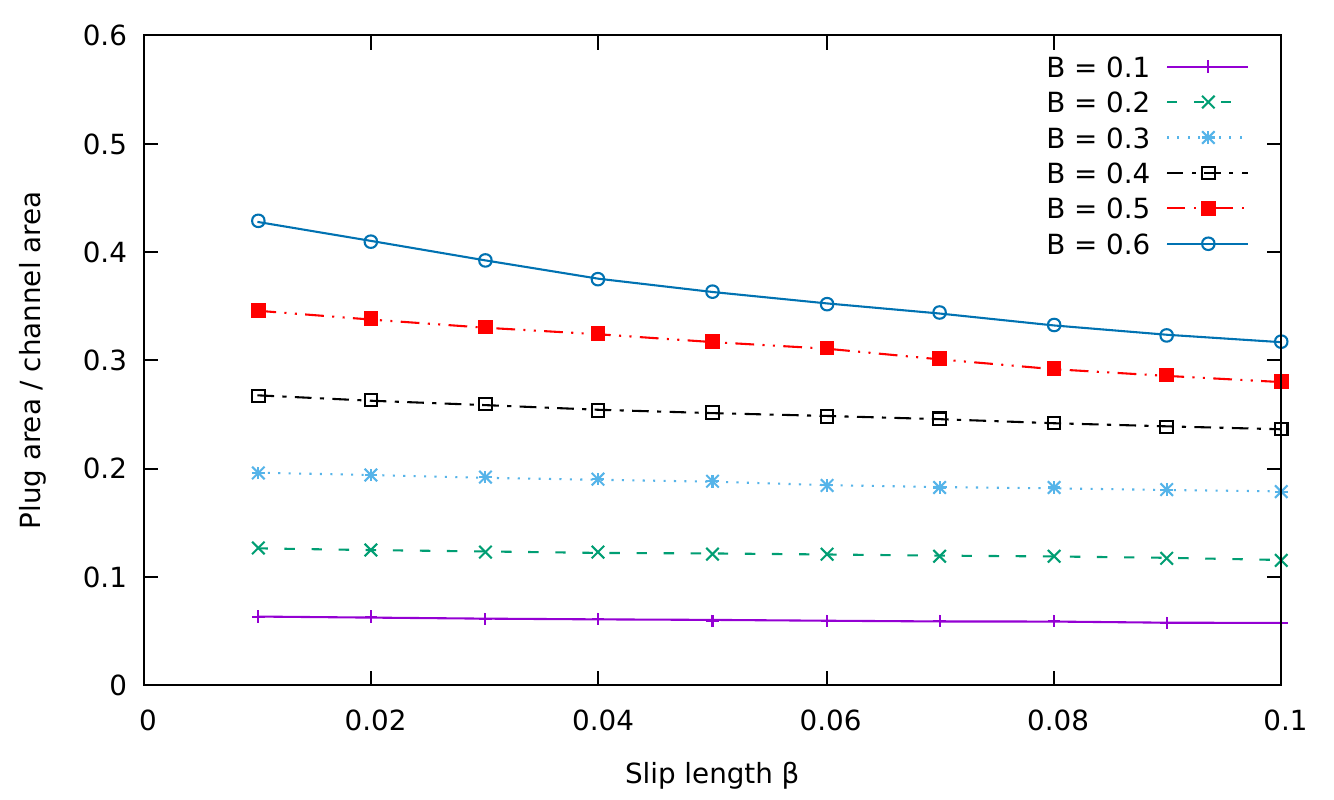}
        \caption{\;}
        \label{fig:plugarea-a}
    \end{subfigure}
    \begin{subfigure}[c]{0.48\textwidth}
        \centering
        \includegraphics[width=\textwidth]{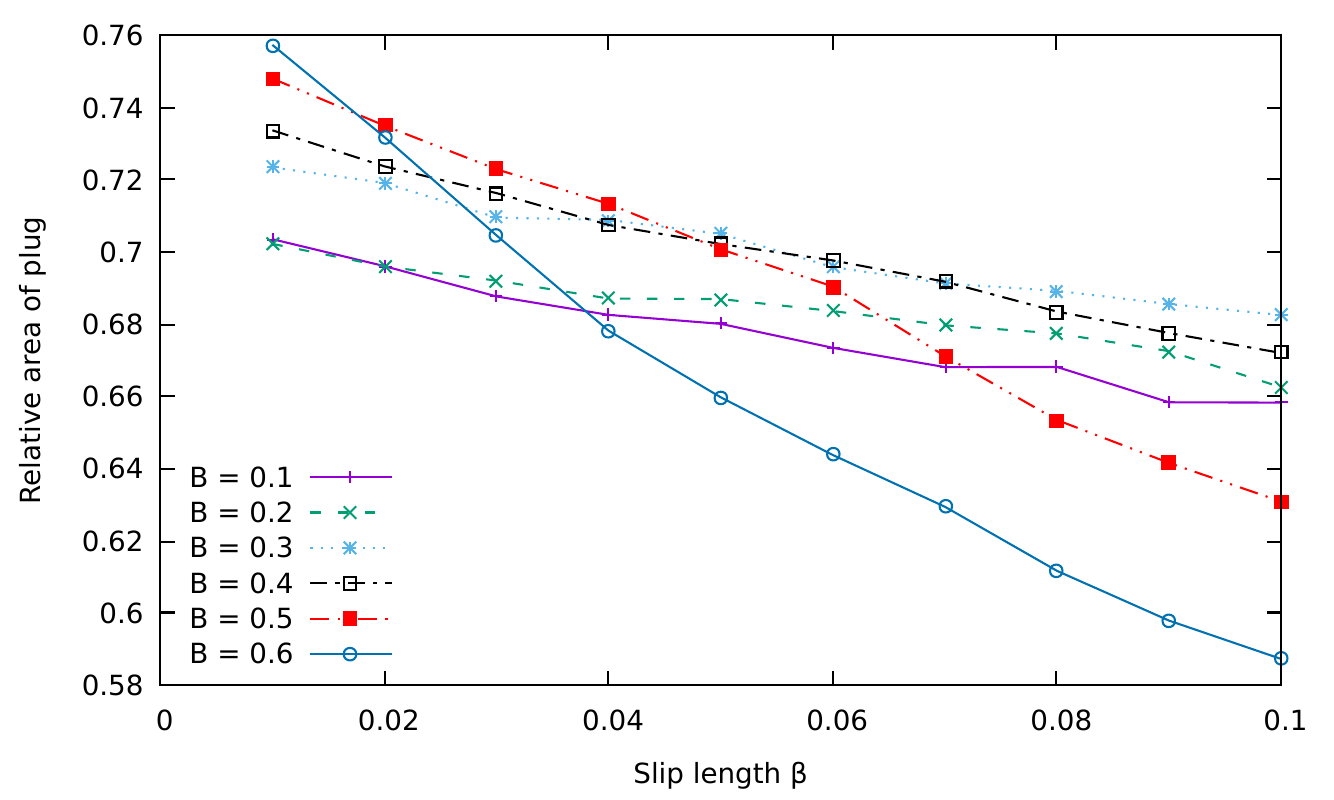}
        \caption{\;}
        \label{fig:plugarea-b}
    \end{subfigure}
\caption{Area of the unyielded plug in the centre of the straight-to-curved channel. (a) Relative to the total area of the channel. In a straight channel the plug area is $B$ times the channel area. (b) Relative to the area of unyielded fluid if there was an abrupt change of plug width from the predicted area in the straight channel to that in the curved part of the channel.}
\label{fig:plugareastraightcurved}
\end{figure}

The length of the transition region in which the plug is suppressed  increases with increasing slip length, and hence the area of the plug region decreases. The plug area relative to the value that would be expected if there was an abrupt change in the plug from the straight channel value (plug area = $LB$) to the curved channel value (area = $\theta_c (\hat{r}_0^2 - \hat{r}_i^2)$) is shown in figure~\ref{fig:plugareastraightcurved}. As the slip length increases the data confirms that the plug gets smaller, more markedly for higher Bingham number. 
 As a proportion of the plug area expected in straight or curved channels, the transition from one channel shape to another leads to a reduction to about 70\% of the predicted area. Although the plug area increases with the Bingham number $B$, its value relative to the expected value is less well behaved. For all $B$ it decreases with slip length: when slip is introduced at the wall the effect of the change of channel curvature is greater for higher slip lengths. Moreover, the variation in relative plug area with slip length is greater for larger $B$. 

\subsection{Sinusoidal channel}
\label{sec:sinusoidal}

\begin{figure}
\centering
    \begin{subfigure}[c]{0.48\textwidth}
        \centering
        \includegraphics[width=\textwidth]{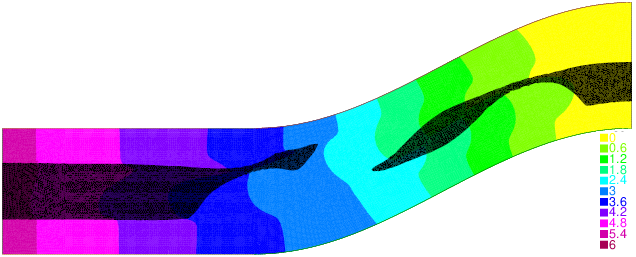}
        \caption{\;}
        \label{fig:comp-yield-sin1a}
    \end{subfigure}
    \begin{subfigure}[c]{0.48\textwidth}
        \centering
        \includegraphics[width=\textwidth]{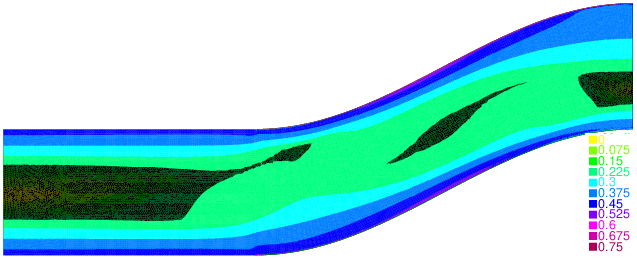}
        \caption{\;}
        \label{fig:comp-yield-sin1b}
    \end{subfigure}

\centering
    \begin{subfigure}[c]{0.48\textwidth}
        \centering
        \includegraphics[width=\textwidth]{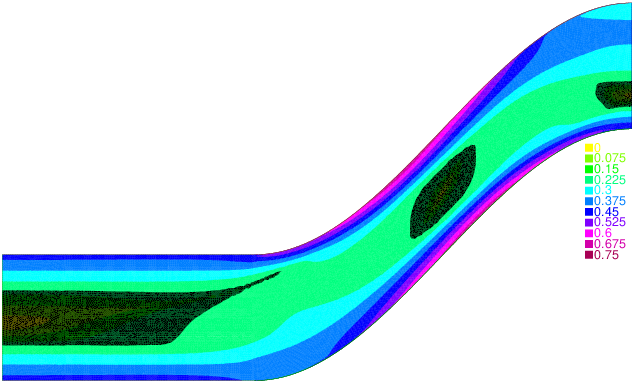}
        \caption{\;}
        \label{fig:comp-yield-sin1c}
    \end{subfigure}
    \begin{subfigure}[c]{0.48\textwidth}
        \centering
        \includegraphics[width=\textwidth]{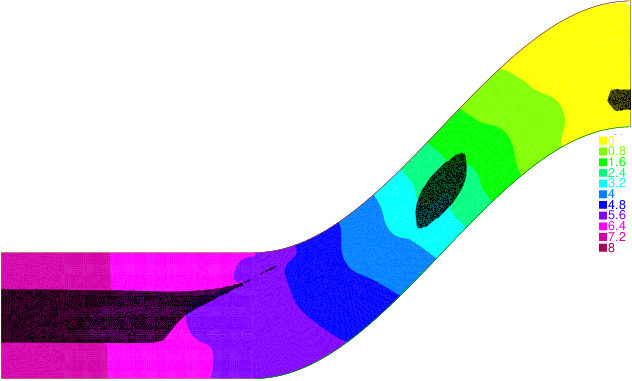}
        \caption{\;}
        \label{fig:comp-yield-sin1d}
    \end{subfigure}

\caption{Numerical solution for a pressure-driven Bingham fluid in a sinusoidal channel, with curved length $L_s = 3$ and Bingham number $B=0.5$. 
(a) Amplitude $y_0 = 0.5$, slip length $\beta=0.02$. The contours shown outside the black unyielded plug are for the pressure. 
(b) Amplitude $y_0 = 0.5$, slip length $\beta=0.1$. Contours of stress. 
(c) Amplitude $y_0 = 1$, slip length $\beta=0.02$. Contours of stress. 
(d) Amplitude $y_0 = 1$, slip length $\beta=0.1$. Contours of pressure. 
}
\label{fig:comp-yield-sin1}
\end{figure}

The second channel geometry that we consider is sinusoidal; examples are shown in figure~\ref{fig:comp-yield-sin1}. The no-slip case is described in~\cite{robertsc23}. In our numerical solutions the channel extends from a straight channel of length $L=2$ into the sinusoidal region. The amplitude of the sine curve is $2y_0$ and its half-length $L_s$. Then the lower wall takes the form
\begin{equation}
 y(x) = y_0 \left(1-\cos \left( \frac{\pi x}{L_s} \right) \right)
\end{equation}
for $0 < x < L_s$ and $y=0$ for $-L < x < 0$.
The upper wall is a distance $W$ higher, although the transverse width of the channel is not quite constant (it narrows slightly, with a weak dependence on $y_0$~\cite{robertsc23}).

\begin{figure}
\centering
    \begin{subfigure}[c]{0.48\textwidth}
        \centering
        \includegraphics[width=\textwidth]{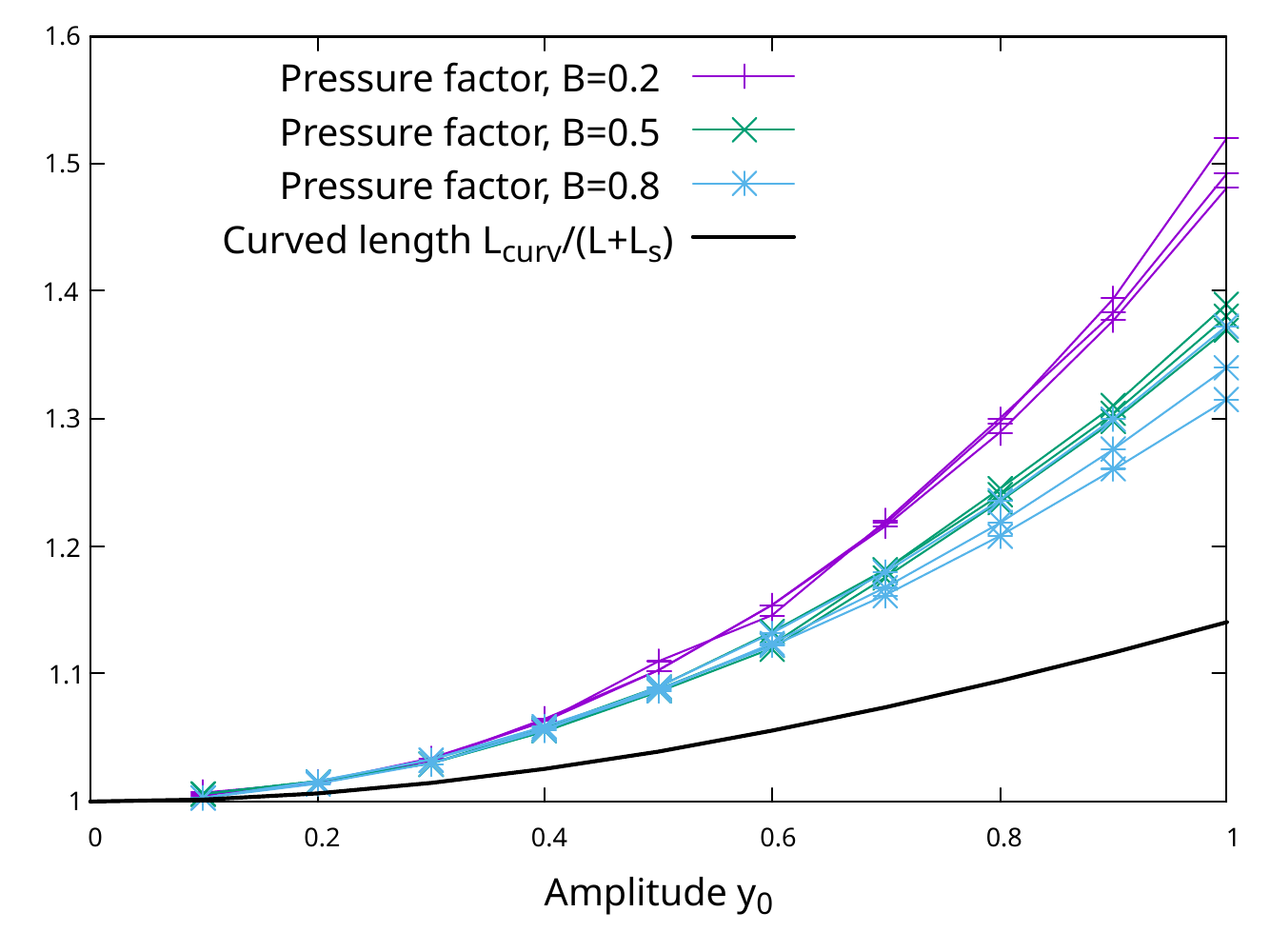}
        \caption{\;}
        \label{fig:pressuresin}
    \end{subfigure}
    \begin{subfigure}[c]{0.48\textwidth}
        \centering
        \includegraphics[width=\textwidth]{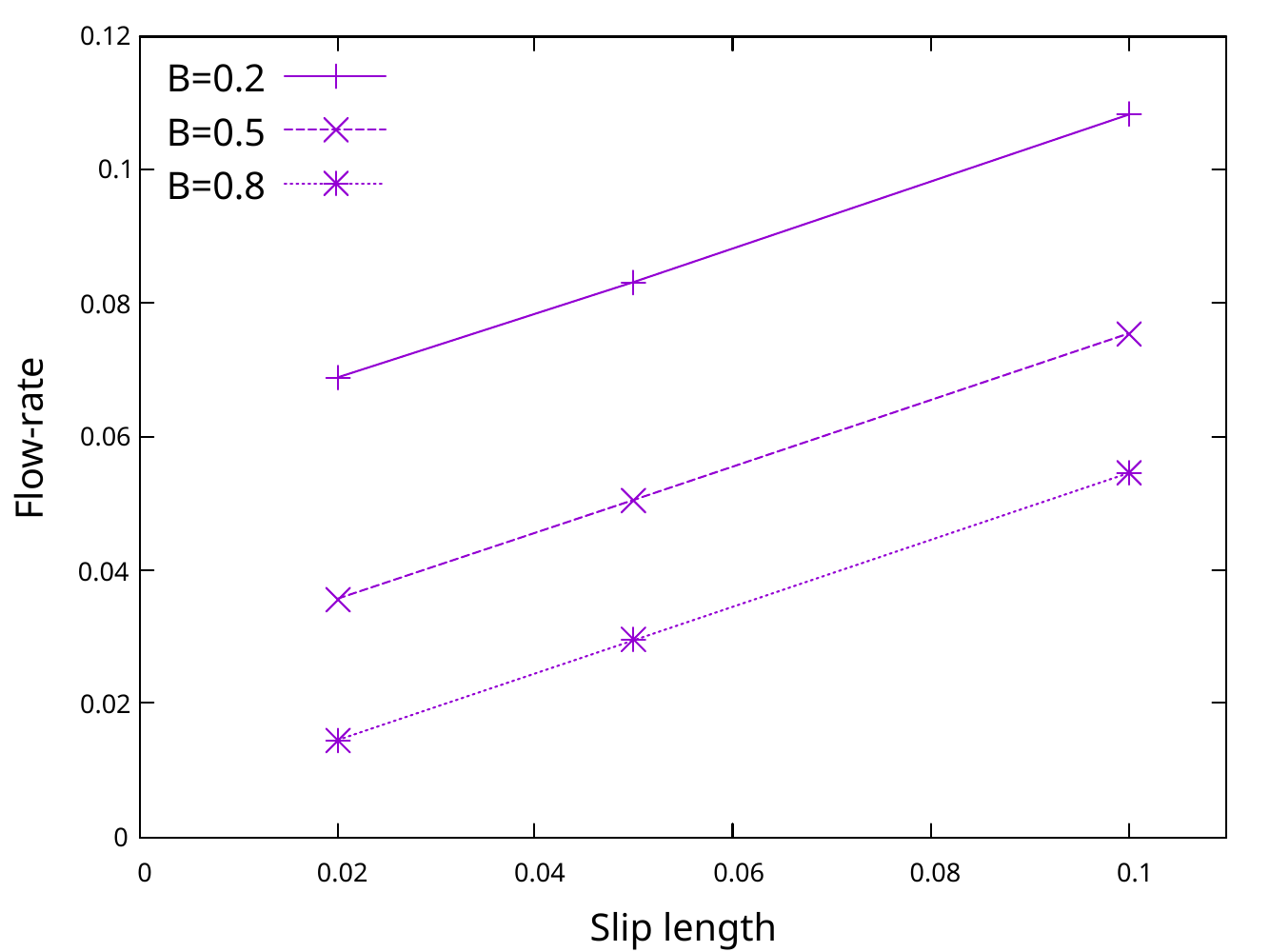}
        \caption{\;}
        \label{fig:sin-flowrate}
    \end{subfigure}
\caption{
(a) The solid black line shows the extra length of the centreline of the sinusoidal channel associated with its curvature, in the case $L_s =3$, which increases with amplitude.
There is a further correction factor to the inlet pressure required to maintain a unit pressure gradient due to the narrowing of the channel, and its effect on the flow itself, which is found by trial and error. 
For each Bingham number, there are three lines shown for slip lengths $\beta = 0.02, 0.05$ and $0.1$, again for $L_s = 3$.
(b) The flow-rates in the sinusoidal channel are independent of the amplitude; they increase linearly with slip-length and decrease with increasing Bingham number.
}
\label{fig:pressuresinQ}
\end{figure}

The numerical details are as for the straight-to-curved channel, except that because of the slight narrowing of the central section of the sinusoidal channel the required inlet pressure 
is unknown. We first calculate the arc length $L_{curv}$ of the centreline of the channel, which because of the curvature is slightly longer than the straight-line distance between the ends, by numerical integration; we show the result in figure~\ref{fig:pressuresin}. The curved length is up to about 14\% longer for the largest amplitude considered. Secondly, to account for the slight variation in the width of the channel as it curves, we choose the inlet pressure in such a way that the flow is what would be expected in the straight part of the channel (from \S~\ref{sec:straight}) for the given values of $B$ and $\beta$. This introduces a correction factor which we find by trial and error, and which we consider as a multiplier of the curved length of the channel. The results are also shown in  figure~\ref{fig:pressuresin}: the factor increases with channel amplitude, as expected, but it also depends on the Bingham number and, weakly, on the slip length. The correction required is greatest for smaller Bingham numbers and the dependence on slip length is greatest for larger Bingham numbers. The resulting inlet pressures range from $p_{in} = L+L_s$ for zero amplitude, $y_0 =0$, up to about 1.6 times this value for $y_0 = 1$. This means that the flow-rates through the channel are given by eq.~(\ref{eq:flow-rate-straight}), and these are shown in figure~\ref{fig:sin-flowrate}.

With the required pressure gradient selected, we can determine the shape and extent of the unyielded fluid.
Figure~\ref{fig:comp-yield-sin1} shows that a tongue of unyielded fluid extends out from the straight part of the channel, oriented towards the upper wall. The re-positioning towards the outside of the channel seen in the straight-to-curved channel case is absent here, perhaps because the change in curvature is less abrupt. The channel then becomes slightly straighter, and the plug can re-form; this is more likely to happen at lower Bingham number and lower amplitude, and, as for the previous channel geometry, increasing the slip length results in a reduction in the plug area. At the apex of the sinusoid, the pattern of yielding is reminiscent of that in the curved channel, albeit over only a short distance, with a small uniformly curved plug close to the inside of the bend. 

\begin{figure}
\centering
    \begin{subfigure}[c]{0.48\textwidth}
        \centering
        \includegraphics[width=\textwidth]{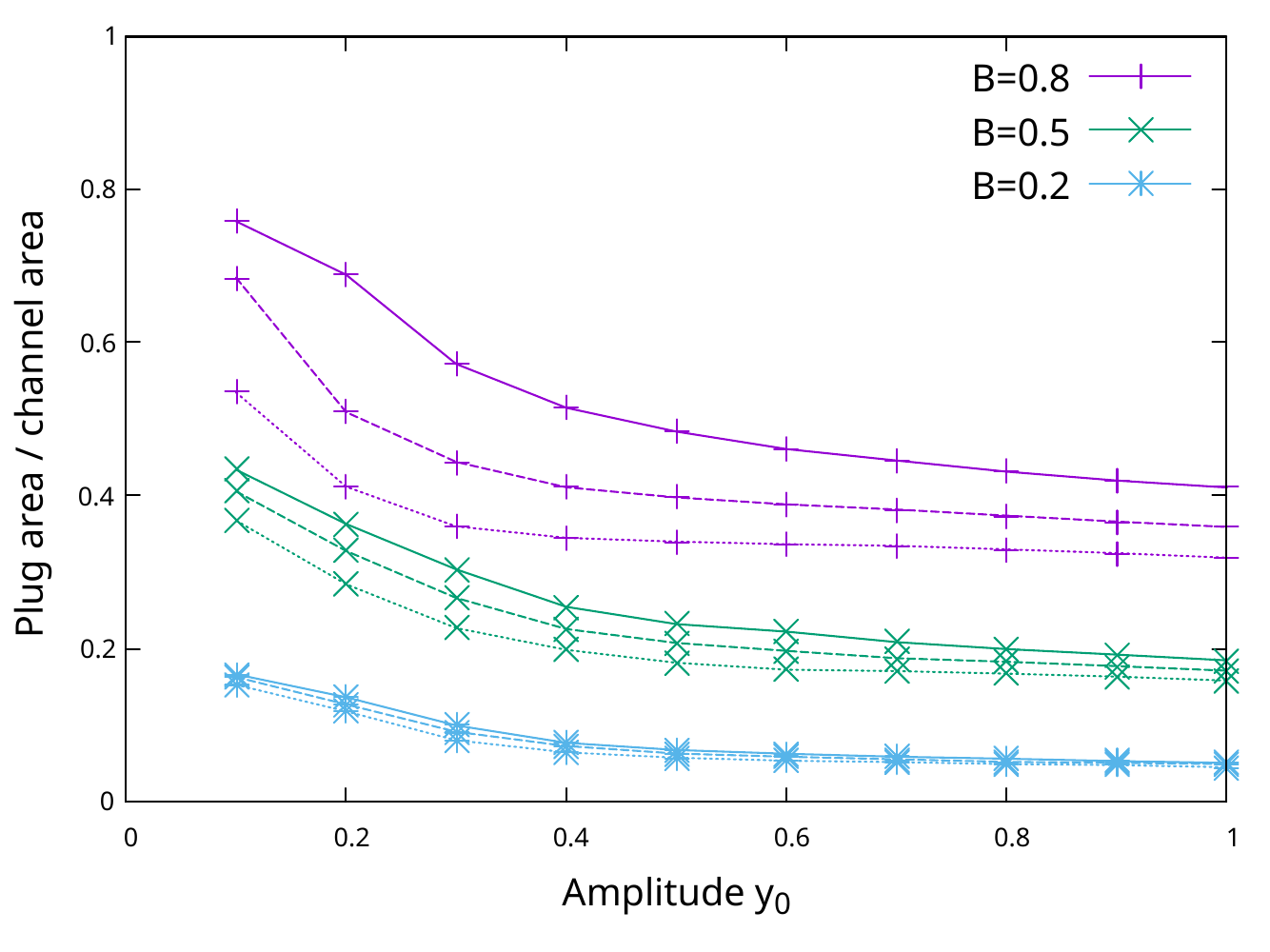}
        \caption{\;}
        \label{fig:plugareasin-a}
    \end{subfigure}
    \begin{subfigure}[c]{0.5\textwidth}
        \centering
        \includegraphics[width=\textwidth]{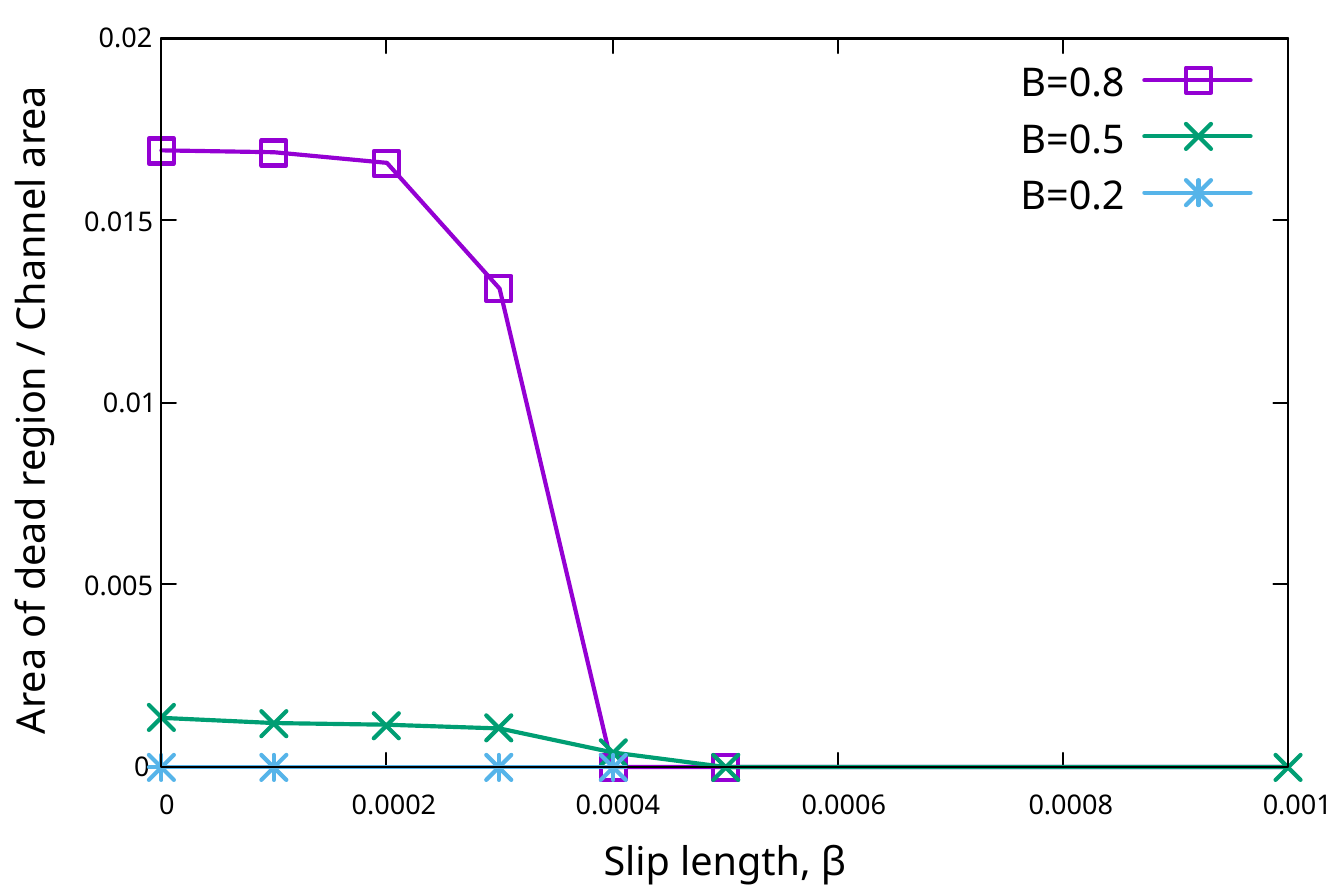}
        \caption{\;}
        \label{fig:plugareasin-b}
    \end{subfigure}
\caption{(a) Area of the unyielded plug in the sinusoidal channel, relative to the total area of the channel (cf. figure~\ref{fig:plugarea-a}).  
Results are shown for three values of the Bingham number, as labelled, and three values of the slip length: $\beta = 0.02$ (dotted lines); $\beta = 0.05$ (dashed lines); $\beta = 0.1$ (solid lines), in a channel with curved length $L_s = 3$.
For zero amplitude the value should be equal to $B$, irrespective of the slip length $\beta$.
(b) Area of the stationary (``dead") unyielded plug at the apex of the sinusoidal channel for small slip lengths. The curved length of the channel is smaller than in (a), $L_s = 2$, and the amplitude is $y_0 = 1$, the largest value studied here. 
}
\label{fig:plugareasin}
\end{figure}

Figure~\ref{fig:plugareasin-a} shows the area of the plug relative to the total channel area. As the channel amplitude increases the plug area decreases, as a consequence of the increased shear stresses in the fluid where it deviates from moving in a straight line. The effect of the slip length is significant for the largest Bingham numbers, and greater at moderate amplitudes. At large amplitude almost all fluid is yielded in the sinusoidal part of the channel, and so the unyielded area should be roughly equal to $BL$, the plug in the straight part of the channel.

At large amplitudes we expect the appearance of dead zones, of stationary unyielded fluid, where the curvature of the channel walls is highest~\cite{robertsc23}. In this case a dead zone is most likely to form at the apex of the channel (for example in the channel with amplitude $y_0 = 1$ shown in figure~\ref{fig:comp-yield-sin1c}) where the stress decreases close to the wall. 
The likelihood of dead zones is reduced by the introduction of slip at the walls. For $L_s = 3$, for example, no dead zones were found for Bingham number up to 0.8 and amplitude up to one, for any slip length.  On reducing the curved length of the channel to $L_s=2$, we show in figure~\ref{fig:plugareasin-b} that the area of the dead zone rapidly decreases with slip length (for $y_0 = 1$).

\section{Discussion and Conclusions}
\label{sec:discussion}

We have derived closed-form expressions for the velocity profile of a Bingham fluid undergoing steady pressure-driven flow through straight and uniformly curved channels in which the fluid can slip at the walls. For a Bingham fluid flowing through a straight channel, we find that the introduction of wall slip increases the flow speed slightly but has no effect on the width of the plug that is formed at the centre of the channel. This is not the case for curved channels. The velocity profile in a curved channel that we derive in \S\ref{sec:curved-deriv} gives a clear indication of both the effects of curvature and slip. The yield surface narrows and moves away from the outer wall. Both the narrowing and the inwards movement of the plug reduce the area of the plug and this is exacerbated by increasing slip length. For sufficiently large slip length and Bingham number the plug region reaches the inner wall.

Our theoretical derivation allows us to validate numerical solutions of the equations of flow in non-uniformly curved channels with more complex geometries, which may be better representations of foam flow in veins. The Bingham number $B$ is dependent on the bubble size and the liquid fraction of the foam~\cite{robertsclj21}. Reducing the liquid fraction, for example, will increase $B$, widening the plug. We chose to set $B$ using the pressure gradient rather than, say, the flow-rate. As a result, increasing the slip-length at fixed $B$ results in faster flows and, in a curved channel, narrower plug regions.  We find that the reduction in the plug size as a consequence of changes in channel wall curvature~\cite{robertsc20} are amplified by slip at the walls.  We also quantify the rapidity with which dead zones in the most curved parts of a channel, in our case the apex of the bend in the sinusoidal channel,  are eliminated with even a small amount of slip.

In the treatment of varicose veins with foam sclerotherapy, an aqueous foam is injected into the dysfunctional vein to displace blood and deliver sclerosant to the vein walls. To do so effectively requires that a plug of foam is formed in the vein and can move along the vein to perform the displacement, and that a high proportion of the cross-section of the vein is swept. Any curvature of the vein walls and any slip of the foam there will affect the efficacy of the process. 
The narrowing of the plug in a curved vein reduces the proportion of the vein that is well-swept.  If the inner boundary of the plug region reaches the inner wall then the displacement effectively ceases.  Our results underline the benefits of keeping the vein as straight as possible when performing sclerotherapy.

The presence of slip narrows the plug even further.  The variation in relative plug area with slip length is greater for larger Bingham numbers, which is the regime in which sclerotherapy aims to operate (with large plug widths) and thus wall slip could cause a significant reduction in the efficacy of the process. Hence if a high degree of wall slip in the vein is expected, the liquid fraction or bubble size of the sclerosing foam should be reduced, or the applied pressure gradient increased, i.e. the pressure applied to a syringe to inject the foam. On the other hand, slip eliminates dead zones in highly curved veins. In the context of foam sclerotherapy, although they may help in delivering sclerosant to the vein walls, dead zones do not contribute to displacing the blood from the vein and should therefore be avoided.  
One could consider other forms of the slip law, e.g. stick-slip laws~\cite{Chaparian_Tammisola_2021}. These are likely to change the results quantitatively, but perhaps not qualitatively: the introduction of slip reduces the plug area in the flow and eliminates dead zones whatever the functional form used to model it. 

Delivery of the sclerosant-containing foam occurs over tens to hundreds of seconds, short enough that the foam doesn't significantly degrade due to gravity-driven liquid drainage, film rupture, or gas diffusion between bubbles with different gas pressures. It is therefore possible that there could be a significant transient at both the beginning and the end of the treatment, in which our steady-state results are less relevant. Further work could determine the importance of a ramp in the pressure gradient (or flow rate) up to, or down from, its steady value. 
Further, it might then be possible to predict motion of the foam along the vein and the dynamics of the displacement of the blood. One could also include the effect of gravity~\cite{taghavi18}, and ask how bouyancy affects the displacement flow in the presence of slip. Experimental validation of our predictions might be tackled in biomimetic veins~\cite{carugo}, where the foam flow could be tracked more easily.

\section*{Acknowledgements}

We acknowledge funding from the Welsh Government and the Government of Qu{\'e}bec through the Wales-Qu{\'e}bec cooperation project {(grant number GQ139359)}. We are grateful to T.G. Roberts and A. Joulaei for discussions and advice.

\end{document}